\documentclass[prd,showpacs,amsmath,amssymb,twocolumn]{revtex4}

\usepackage{graphicx}
\usepackage{dcolumn}
\usepackage{bm}
\usepackage{psfrag}
\usepackage{subfigure}

\newcommand{\be}{\begin{eqnarray}}
\newcommand{\ee}{\end{eqnarray}}
\newcommand{\bn}{\begin{eqnarray*}}
\newcommand{\en}{\end{eqnarray*}}

\newcommand{\nn}{\nonumber \\}
\newcommand{\nl}{\\}

\renewcommand{\d}{\mbox{\rm d}}

\renewcommand{\th}{\ensuremath{\theta}}

\newcommand{\ph}{\ensuremath{\phi}}

\newcommand{\al}{\ensuremath{\alpha}}
\newcommand{\bt}{\ensuremath{\beta}}
\newcommand{\sg}{\ensuremath{\sigma}}
\newcommand{\gm}{\ensuremath{\gamma}}
\newcommand{\dl}{\ensuremath{\delta}}
\newcommand{\lm}{\ensuremath{\lambda}}

\newcommand{\Dl}{\ensuremath{\Delta}}
\newcommand{\Sg}{\ensuremath{\Sigma}}
\newcommand{\Gm}{\ensuremath{\Gamma}}

\newcommand{\OmK}{\ensuremath{\Omega_{\rm K}}}

\newcommand{\Cmu}{\ensuremath{\hat{\mu}}}
\newcommand{\Cnu}{\ensuremath{\hat{\nu}}}
\newcommand{\Cal}{\ensuremath{\hat{\al}}}
\newcommand{\Cbt}{\ensuremath{\hat{\bt}}}

\newcommand{\Ci}{\ensuremath{\hat{\imath}}}

\newcommand{\ze}{\ensuremath{\hat{0}}}
\newcommand{\on}{\ensuremath{\hat{1}}}
\newcommand{\tw}{\ensuremath{\hat{2}}}
\newcommand{\tr}{\ensuremath{\hat{3}}}

\newcommand{\lt}{\ensuremath{\left}}
\newcommand{\rt}{\ensuremath{\right}}

\renewcommand{\d}{\mbox{\rm d}}

\begin{document}

\pagenumbering{arabic}

\title{Perturbation Method for Classical Spinning Particle Motion:  I. Kerr Space-Time}

\author{Dinesh Singh}
\email{dinesh.singh@uregina.ca}
\affiliation{%
Department of Physics, University of Regina \\
Regina, Saskatchewan, S4S 0A2, Canada
}%
\date{\today}

\begin{abstract}
This paper presents an analytic perturbation approach to the dynamics of a classical spinning particle,
according to the Mathisson-Papapetrou-Dixon (MPD) equations of motion, with a direct application to circular motion
around a Kerr black hole.
The formalism is established in terms of a power series expansion with respect to the particle's spin magnitude,
where the particle's kinematic and dynamical degrees are expressed in a completely general form that can be constructed
to infinite order in the expansion parameter.
It is further shown that the particle's squared mass and spin magnitude can shift due to a classical analogue of radiative corrections
that arise from spin-curvature coupling.
Explicit expressions are determined for the case of circular motion near the event horizon a Kerr black hole,
where the mass and spin shift contributions are dependent on the initial conditions of the particle's spin orientation.
A preliminary analysis of the stability properties of the orbital motion in the Kerr background due to spin-curvature interactions
is explored and briefly discussed.
\end{abstract}


\maketitle

\section{Introduction}
\label{sec:1}

One of the earliest and on-going research interests in general relativity concerns the dynamics of extended bodies in
the presence of strong gravitational backgrounds.
Considering that virtually all astrophysical objects in the Universe, such as black holes, neutron stars, and other
isolated massive bodies, have at least some spin angular momentum in their formation, it is not difficult to surmise
that an in-depth study of moving relativistic systems with spin is a useful endeavour.
A relevant example concerns the motion of rapidly rotating neutron stars in circular orbit around supermassive black holes
like ones believed to exist in the centre of galaxies, which serve as candidate sources for emitting low-frequency gravitational
wave radiation that may be detected by the space-based LISA gravitational wave observatory \cite{LISA}.

A first attempt to understand the dynamics of extended bodies in curved space-time was put forward by Mathisson \cite{Mathisson},
who showed the existence of an interaction term involving the direct coupling of particle spin to the Riemann curvature tensor
generated by a background source.
Steady progress was made since this first attempt, with a notable contribution made several years afterwards
by Papapetrou \cite{Papapetrou}, who proposed that the spinning particle exists within a space-time world tube
containing its centre-of-mass worldline, where its associated matter field has compact support.
In addition, multipole moment contributions, i.e. beyond the mass monopole and spin dipole, to the extended objects
full equations of motion were considered by Tulczyjew \cite{Tulczyjew} and others, ultimately leading to the
expressions obtained by Dixon \cite{Dixon1,Dixon2}, with a self-consistent description for all multipole moment
contributions to infinite order.
While the various theories of extended body motion in curved space-time differ with respect to the
higher-order multipole moments, all of them recover the ``pole-dipole approximation'' identified initially by
Mathisson and Papapetrou, which are satisfactory for most practical calculations, so long as the
dimensions of the spinning body are small when compared to the background space-time's local radius of curvature.
These truncated expressions of the full equations of motion are commonly known as the Mathisson-Papapetrou-Dixon (MPD) equations.

There has been widespread interest in applying the MPD equations to the dynamics of classical spinning particles
in orbit around rotating black holes, as described by the Kerr metric \cite{Mashhoon1,Wald,Tod,Semerak,Suzuki1}.
In many ways, the Kerr background is an ideal testing ground for the MPD equations, since both mass sources
are spinning, which introduce interesting spin-curvature effects that impact upon the orbiting particle's overall evolution.
Furthermore, it lends itself well to numerical simulations of deterministic chaos under extreme conditions \cite{Suzuki1,Suzuki2,Hartl1,Hartl2},
as well as studies of gravitational wave generation \cite{Mino,Tanaka} arising from spin-induced deviations away from geodesic motion.

More formal study of the MPD equations have also occurred in various forms \cite{Ehlers,Bailey,Noonan},
including a recent perturbative approach developed by Chicone, Mashhoon, and Punsly (CMP) \cite{Chicone},
with application to the study of rotating plasma clumps propagating in astrophysical jets directed along a Kerr black hole's
axis of symmetry.
Another application of the CMP approximation by Mashhoon and Singh \cite{Mashhoon2} determined analytic expressions for
leading-order spin-curvature perturbations of a spinning particle's circular orbit around a Kerr black hole.
This analysis is successful in reproducing the spinning particle's kinematic behaviour compared to numerical
simulations of the full MPD equations for situations where, for spin magnitude $s$ and mass $m$, the M{\o}ller radius \cite{Mashhoon2,Moller}
for the spinning particle is $s/m \lesssim 10^{-3} \, r$, and $r$ is the particle's radial distance away from the background mass source.
However, this approximation starts to break down when $s/(m r) \sim 10^{-2}-10^{-1}$ for $r = 10 \, M$,
where $M$ is the Kerr black hole mass, suggesting that higher-order spin-curvature coupling terms are required to more
completely describe the orbital motion.

It was for this initial purpose that a generalization of the CMP approximation was very recently introduced by Singh \cite{Singh0}
to incorporate higher-order analytic contributions to the perturbation approach for the MPD equations.
This generalization has several nice features.
For example, as a power series expansion with respect to the particle's spin magnitude,
it can be extended to formally {\em infinite order} in the expansion.
In addition, it leads to expressions that are background independent, and is fully applicable to {\em arbitrary motion} of
the particle, without recourse to any space-time symmetries within the metric.
As a result, this generalization is very robust, with applicability for many distinct scenarios in theoretical astrophysics,
such as the modelling of globular clusters and other many-body dynamical systems in curved space-time, and also spinning particle
interactions with gravitational waves, the results of which can be compared with existing treatments \cite{Nieto,Mohseni,Kessari}.
Furthermore, this approach identifies the existence of a classical analogue for ``radiative corrections''
that shift the particle's overall squared mass and spin magnitude due to higher-order spin-curvature contributions,
a feature not thought about before.
It would, therefore, be very useful to investigate the computational capacity of this generalization when
applied to circular motion in the Kerr background.
This is especially so in extreme conditions where a transition from stable to chaotic motion may be analytically identified,
for comparison with existing approaches \cite{Suzuki1,Suzuki2,Hartl1,Hartl2} which use primarily numerical methods.

The purpose of this paper is to present the generalized form of the CMP approximation for the MPD equations within the context
of circular motion around a Kerr black hole, and explore the derived physical consequences.
It begins with Sec.~\ref{sec:2}, which displays the full MPD equations, followed by a presentation of the formalism
behind the generalized CMP approximation in Sec.~\ref{sec:3}.
Afterwards, Sec.~\ref{sec:4} presents the formal application of the generalized CMP approximation to the case
of circular motion around a Kerr black hole, up to second-order in the perturbation expansion parameter.
This is followed, in Sec.~\ref{sec:5}, by analysis of the predicted kinematic and dynamical properties of the perturbed system,
including the predicted effective squared mass and spin magnitude of the spinning particle.
A general discussion of the main results obtained in this paper is found in Sec.~\ref{sec:6}, with a brief conclusion thereafter.
The metric convention adopted is $+2$ signature with Riemann and Ricci tensor definitions following MTW \cite{MTW}, and
geometric units of $G = c =1$ are assumed throughout.

\section{Mathisson-Papapetrou-Dixon (MPD) Equations}
\label{sec:2}

The MPD equations of motion for the spinning particle's linear four-momentum $P^\mu(\tau)$ and spin tensor
$S^{\al \bt}(\tau)$ consist of
\begin{subequations}
\label{MPD-equations}
\be
{DP^\mu \over \d \tau} & = & - {1 \over 2} \, R^\mu{}_{\nu \al \bt} \, u^\nu \, S^{\al \bt} \, ,
\label{MPD-momentum}
\nl
\nn
{DS^{\al \bt} \over \d \tau} & = & P^\al \, u^\bt - P^\bt \, u^\al \, ,
\label{MPD-spin}
\ee
\end{subequations}
where (\ref{MPD-momentum}) describes the force applied due to spin-curvature coupling via $S^{\al \bt}(\tau)$, the particle's
four-velocity vector $u^\mu(\tau) = \d x^\mu(\tau)/\d \tau$ with affine parametrization $\tau$,
and the Riemann curvature tensor $R_{\mu \nu \al \bt}$, while (\ref{MPD-spin}) describes the corresponding torque generated.
As a result of (\ref{MPD-spin}), the particle's four-momentum precesses around the centre-of-mass worldline.
While it is possible to identify $\tau$ with proper time such that $u^\mu \, u_\mu = -1$, it is useful to leave it
unspecified at present.
The differences between competing descriptions of extended objects in curved space-time arise from differing higher-order
multipole moment terms beyond the mass monopole and spin dipole moment, leading to additive contributions of the form
$\cal F^\mu$ and $\cal T^{\al \bt}$ \cite{Ehlers,Mashhoon2} in (\ref{MPD-momentum}) and (\ref{MPD-spin}), respectively.
Specification of $\cal F^\mu$ and $\cal T^{\al \bt}$ requires knowledge of the spinning object's energy-momentum tensor $T^{\mu \nu}$
\cite{Dixon1,Dixon2,Mashhoon1,Mashhoon2}, satisfying covariant conservation $T^{\mu \nu}{}_{; \nu} = 0$.
For most practical purposes, however, the expression for (\ref{MPD-equations}) is sufficient.

At present, the MPD equations as expressed in (\ref{MPD-equations}) are underdetermined, and require
additional equations to completely specify the system.
Following the approach from Dixon \cite{Dixon1,Dixon2}, the orthogonality spin condition
relating the particle's linear and spin angular momenta according to
\be
S^{\alpha \beta} \, P_\beta & = & 0 \,
\label{spin-condition}
\ee
is introduced, while the mass and spin parameters $m$ and $s$ naturally take the form
\begin{subequations}
\label{m-s-magnitude}
\be
m^2 & = & -P_\mu \, P^\mu \, ,
\label{mass}
\nl
\nn
s^2 & = & {1 \over 2} \, S_{\mu \nu} \, S^{\mu \nu} \, .
\label{spin}
\ee
\end{subequations}
It is very important to note that, while (\ref{mass}) and (\ref{spin}) are both technically functions of $\tau$,
the MPD equations and (\ref{spin-condition}) indicate that $m$ and $s$ are {\em constants of the motion} \cite{Chicone}.
As well, the combination of (\ref{MPD-equations}) and (\ref{spin-condition}) are known \cite{Tod} to result
in an expression for the four-velocity $u^\mu$ in terms of $P^\mu$ and $S^{\al \bt}$, such that
\be
u^\mu & = & -{P \cdot u \over m^2} \lt[P^\mu
+ {1 \over 2} \, {S^{\mu \nu} \, R_{\nu \gm \al \bt} \, P^\gm \, S^{\al \bt} \over
m^2 + {1 \over 4} \, R_{\al \bt \rho \sg} \, S^{\al \bt} \, S^{\rho \sg}} \rt],
\label{MPD-velocity}
\ee
where $P \cdot u$ is currently an undetermined scalar product relating the particle's internal clock with respect to $\tau$.
It becomes a self-evident confirmation from (\ref{MPD-velocity}) that spin-curvature coupling displaces the particle's
four-velocity away from a geodesic in curved space-time, leading to a dynamically rich interplay between the
particle's centre-of-mass motion and its dynamical response due to spin-curvature interaction.
The constraint equations (\ref{spin-condition})--(\ref{MPD-velocity}) will prove very useful for ultimately deriving the
generalized CMP approximation \cite{Singh0}.

\section{The Generalized CMP Approximation for the MPD Equations}
\label{sec:3}

\subsection{CMP Approximation}
\label{sec:3.1}

The approach taken by Chicone, Mashhoon, and Punsly in deriving the CMP approximation \cite{Chicone,Mashhoon2} is to first
assume that $P^\mu - m \, u^\mu = E^\mu$ is a small quantity, where $E^\mu$ is the spin-curvature force.
As well, the M{\o}ller radius $\rho$ \cite{Chicone,Mashhoon2,Moller} is chosen to be small, such that $\rho = s/m \ll r$,
where $r$ is the distance from the particle to the background gravitational source.
This combination leads to the CMP approximation for the MPD equations, a series expansion to first-order in $s$, such that
\begin{subequations}
\label{CMP-approximation}
\be
{D P^\mu \over \d \tau} & \approx & -{1 \over 2} \, R^\mu{}_{\nu \al \bt} \, u^\nu \, S^{\al \bt}\, ,
\label{CMP-momentum}
\nl
\nn
{DS^{\al \bt} \over \d \tau} & \approx & 0 \, ,
\label{CMP-spin}
\ee
\end{subequations}
where the spin tensor in (\ref{CMP-spin}) is parallel transported within the approximation, and the spin condition
(\ref{spin-condition}) takes the form
\be
S^{\al \bt} \, u_\bt & \approx & 0 \, ,
\label{CMP-spin-condition}
\ee
coinciding with the Pirani condition \cite{Pirani} relating the orthogonality of the spin tensor to the four-velocity.

The CMP approximation is a useful first step in an analytic perturbation approach to the MPD equations,
with a remarkably accurate description for circular motion around a Kerr black hole \cite{Mashhoon2} compared to
the full MPD equations for $s/(mr) \sim 10^{-3}$ and $r = 10 \, M$.
However, it becomes clear that the CMP approximation breaks down as $s/(mr) \sim 10^{-2}-10^{-1}$ for the same choice of $r$,
which follows from the loss of torque information due to (\ref{CMP-spin}), especially since the spin-induced modulation
of the particle's $\tau$-dependent radial position found in the MPD equations is not present in the CMP approximation.
This weakness within (\ref{CMP-approximation}) and (\ref{CMP-spin-condition}) is suggestive of a more detailed and systematic
approach that has resulted in the generalization to follow \cite{Singh0}.

\subsection{Generalization of the CMP Approximation}
\label{sec:3.2}

\subsubsection{Formalism}
The approach taken to generalize the CMP approximation is to assume a power series expansion of the particle's linear momentum and
spin angular momentum, such that
\begin{subequations}
\label{P-S-approx}
\be
P^\mu(\varepsilon) & \equiv & \sum_{j = 0}^\infty \varepsilon^j \, P_{(j)}^\mu \, , 
\label{P-approx-def}
\nl
\nn
S^{\mu \nu} (\varepsilon) & \equiv & \varepsilon \sum_{j = 0}^\infty \varepsilon^j \, S_{(j)}^{\mu \nu}
\ = \ \sum_{j = 1}^\infty \varepsilon^j \, S_{(j-1)}^{\mu \nu} \, ,
\label{S-approx-def}
\ee
\end{subequations}
where $\varepsilon$ is an expansion parameter to be associated with $s$, and
$P_{(j)}^\mu$ and $S_{(j-1)}^{\mu \nu}$ are the respective jth-order contributions of the linear momentum and spin angular momentum in $\varepsilon$.
This implies that the zeroth-order expressions in $\varepsilon$ denote the dynamics of a spinless particle in geodesic motion.
As well, the four-velocity is assumed to take the form
\be
u^\mu(\varepsilon) & \equiv & \sum_{j = 0}^\infty \varepsilon^j \, u_{(j)}^\mu \, .
\label{MPD-velocity-e0}
\ee
When substituting (\ref{P-S-approx}) and (\ref{MPD-velocity-e0}) into the MPD equations described by
\begin{subequations}
\label{MPD-e}
\be
{DP^\mu (\varepsilon) \over \d \tau} & = & -{1 \over 2} \, R^\mu{}_{\nu \al \bt} \, u^\nu (\varepsilon) \, S^{\al \bt} (\varepsilon) \, ,
\label{MPD-momentum-e}
\nl
\nn
{DS^{\al \bt}(\varepsilon) \over \d \tau} & = &  2 \, \varepsilon \, P^{[\al}(\varepsilon) \, u^{\bt]}(\varepsilon) \, ,
\label{MPD-spin-e}
\ee
\end{subequations}
where an extra factor of $\varepsilon$ is introduced in (\ref{MPD-spin-e}) for consistency,
it follows that the jth-order expressions of the MPD equations are
\begin{subequations}
\label{MPD-j}
\be
{DP_{(j)}^\mu \over \d \tau} & = & - \frac{1}{2} \, R^\mu{}_{\nu \alpha \beta} \sum_{k = 0}^{j - 1} u_{(j-1-k)}^\nu \, S_{(k)}^{\alpha \beta} \, ,
\label{DP-j}
\nl
\nn
{DS_{(j-1)}^{\alpha \beta} \over \d \tau} & = &  2 \sum_{k = 0}^{j-1} P_{(j-1-k)}^{[\alpha} \, u_{(k)}^{\beta]} \, .
\label{DS-j}
\ee
\end{subequations}
Given $P_{(0)}^\mu = m_0 \, u_{(0)}^\mu$, where
\be
m_0^2 & \equiv & -P^{(0)}_\mu \, P_{(0)}^\mu \, ,
\label{m0-sq}
\ee
it can be shown that the zeroth-order term in $\varepsilon$ is
\be
{D P_{(0)}^\mu \over \d \tau} & = & 0 \, ,
\label{DP-0}
\ee
while the respective first-order terms following (\ref{MPD-j}) are
\begin{subequations}
\label{MPD-1}
\be
{D P_{(1)}^\mu \over \d \tau} & = & -{1 \over 2} \, R^\mu{}_{\nu \al \bt} \, u_{(0)}^\nu \, S_{(0)}^{\al \bt} \, ,
\label{DP-1}
\nl
\nn
{D S_{(0)}^{\al \bt} \over \d \tau} & = & 0 \, ,
\label{DS-0}
\ee
\end{subequations}
which is the CMP approximation.

\subsubsection{Supplementary Equations}

A complete specification of (\ref{MPD-j}) requires determining $u_{(j)}^\mu$ as a function of the linear and spin angular momentum
expansion components.
In turn, this requires use of the supplementary equations (\ref{spin-condition})--(\ref{MPD-velocity}),
which have important consequences within the formalism.
It is straightforward to show that the spin condition (\ref{spin-condition}) in terms of (\ref{P-approx-def}) and (\ref{S-approx-def}) is
\be
P_\mu^{(0)} \, S_{(j)}^{\mu \nu} & = & - \sum_{k=1}^j P_\mu^{(k)} \, S_{(j-k)}^{\mu \nu} \, , \qquad j \geq 1
\label{s.p=0}
\ee
for the (j+1)th-order contribution in $\varepsilon$, where the first-order perturbation in $\varepsilon$ is
\be
P_\mu^{(0)} \, S_{(0)}^{\mu \nu} & = & 0 \, .
\label{s.p=0-e1}
\ee
Given that the mass and spin magnitude parameters $m$ and $s$ described by (\ref{m-s-magnitude}) are dependent on
$P^\mu$ and $S^{\mu \nu}$, which are represented by (\ref{P-approx-def}) and (\ref{S-approx-def}), respectively,
it is possible to identify a classical analogue of a {\em bare mass} $m_0$ defined by (\ref{m0-sq}) and a {\em bare spin} $s_0$,
according to
\be
s_0^2 & \equiv & {1 \over 2} \, S_{\mu \nu}^{(0)} \, S_{(0)}^{\mu \nu} \, ,
\label{s0-sq}
\ee
in analogy with the radiative corrections identified with the bare mass and spin parameters in quantum field theory.
In this way, the MPD equations in perturbative form yield total mass and spin magnitudes due to the sum of
``radiative corrections'' to $m_0$ and $s_0$, such that
\be
m^2 (\varepsilon) & = & m_0^2 \lt(1 + \sum_{j=1}^\infty \varepsilon^j \, \bar{m}_j^2 \rt),
\label{m-sq}
\nl
\nn
s^2 (\varepsilon) & = & \varepsilon^2 \, s_0^2 \lt(1 + \sum_{j=1}^\infty \varepsilon^j \, \bar{s}_j^2 \rt),
\label{s-sq}
\ee
where
\be
\bar{m}_j^2 & = & - {1 \over m_0^2} \, \sum_{k=0}^j P_\mu^{(j-k)} \, P_{(k)}^\mu \, ,
\label{m-bar-sq}
\nl
\nn
\bar{s}_j^2 & = & {1 \over s_0^2} \, \sum_{k=0}^j S_{\mu \nu}^{(j-k)} \, S_{(k)}^{\mu \nu} \, ,
\label{s-bar-sq}
\ee
are dimensionless jth-order corrections to $m_0^2$ and $s_0^2$, respectively.
Given that $m^2$ and $s^2$ are already shown to be constant within the exact set of MPD equations,
it must also be true that $\bar{m}_j^2$ and $\bar{s}_j^2$ are individually constant for each order of $\varepsilon$.

For the remaining supplementary equation (\ref{MPD-velocity}), the four-velocity described by
\begin{subequations}
\label{MPD-velocity-e}
\be
u^\mu(\varepsilon) 
& = & -{P \cdot u \over m^2(\varepsilon)} \left[P^\mu(\varepsilon) \rt.
\nn
&  &{} + \lt.
{1 \over 2} \, {S^{\mu \nu}(\varepsilon) \, R_{\nu \gamma \alpha \beta} \, P^\gamma(\varepsilon) \,
S^{\alpha \beta}(\varepsilon) \over m^2(\varepsilon) \, \Dl(\varepsilon)} \right],
\label{MPD-velocity-fixed}
\nl
\Dl(\varepsilon) & \equiv & 1 + {1 \over 4 \, m^2(\varepsilon)} \, R_{\mu \nu \al \bt} \, S^{\mu \nu} (\varepsilon) \, S^{\al \bt} (\varepsilon) \, ,
\label{Delta-e}
\ee
\end{subequations}
can be determined as a series expansion in $\varepsilon$, upon specifying the yet undetermined scalar product $P \cdot u$.
With the particularly useful choice of
\be
P \cdot u & \equiv & -m(\varepsilon),
\label{P.u}
\ee
it is straightforward to determine that
\be
u_\mu(\varepsilon) \, u^\mu(\varepsilon) & = & -1 + {1 \over 4 \, m^6(\varepsilon) \, \Dl^2 (\varepsilon)}
\, \tilde{R}_\mu(\varepsilon) \, \tilde{R}^\mu (\varepsilon)
\nn
& = & -1 + O(\varepsilon^4) \, ,
\label{u.u}
\ee
where
\be
\tilde{R}^\mu(\varepsilon) & \equiv & S^{\mu \nu}(\varepsilon) R_{\nu \gm \al \bt} \, P^\gm (\varepsilon) \, S^{\al \bt} (\varepsilon) \, .
\label{R-tilde}
\ee
To at least third-order in $\varepsilon$, the outcome (\ref{u.u}) from (\ref{P.u}) justifies the identification of
$\tau$ as {\em proper time} for parameterization of the particle's centre-of-mass worldline.
It is also straightforward, though tedious, to show explicitly from substituting (\ref{P-S-approx}) and (\ref{m-sq})
into (\ref{MPD-velocity-e}) that the spinning particle's four-velocity in general form is
\begin{widetext}
\be
u^\mu(\varepsilon) & = & \sum_{j = 0}^\infty \varepsilon^j \, u_{(j)}^\mu
\ = \ {P_{(0)}^\mu \over m_0} + \varepsilon \lt[{1 \over m_0} \lt(P_{(1)}^\mu - {1 \over 2} \, \bar{m}_1^2 \, P_{(0)}^\mu\rt) \rt]
\nn
\nn
\nn
&   & {} + \varepsilon^2 \lt\{ {1 \over m_0} \lt[P_{(2)}^\mu - {1 \over 2} \, \bar{m}_1^2 \, P_{(1)}^\mu
         - {1 \over 2} \lt(\bar{m}_2^2 - {3 \over 4} \, \bar{m}_1^4\rt) P_{(0)}^\mu \rt]
+ {1 \over 2 m_0^3} \, S_{(0)}^{\mu \nu} \, R_{\nu \gm \al \bt}  \, P_{(0)}^\gm \, S_{(0)}^{\al \bt} \rt\}
\nn
\nn
\nn
&   & {} + \varepsilon^3 \lt\{ {1 \over m_0} \lt[P_{(3)}^\mu - {1 \over 2} \, \bar{m}_1^2 \, P_{(2)}^\mu
         - {1 \over 2} \lt(\bar{m}_2^2 - {3 \over 4} \, \bar{m}_1^4\rt) P_{(1)}^\mu \rt.
- {1 \over 2} \lt(\bar{m}_3^2 - {3 \over 2} \, \bar{m}_1^2 \, \bar{m}_2^2 + {5 \over 8} \, \bar{m}_1^6 \rt) P_{(0)}^\mu \rt]
\nn
\nn
&   & {} + \lt.
{1 \over 2 m_0^3} \, R_{\nu \gm \al \bt} \lt[\sum_{n=0}^1 S_{(1-n)}^{\mu \nu} \sum_{k=0}^n P_{(n-k)}^\gm \, S_{(k)}^{\al \bt}
- {3 \over 2} \, \bar{m}_1^2 \, S_{(0)}^{\mu \nu} \, P_{(0)}^\gm \, S_{(0)}^{\al \bt} \rt] \rt\}
+ O(\varepsilon^4) \, ,
\label{MPD-velocity-explicit}
\ee
\end{widetext}
which also satisfies (\ref{u.u}) to third order in $\varepsilon$.
Following (\ref{MPD-velocity-explicit}), it can be verified that \cite{Singh0}
\be
{D \bar{s}_j^2 \over \d \tau} & = & {D \bar{m}_j^2 \over \d \tau} \ = \ 0 \,
\label{Ds/dt-Dm/dt=0}
\ee
up to third-order in $\varepsilon$, with the use of (\ref{MPD-j}).

\subsubsection{Perturbations of the M{\o}ller Radius}

A useful consideration within the generalized CMP approximation is the perturbation expression for the M{\o}ller radius $\rho = s/m$,
since this quantity is relevant for identifying the strength of the spin-curvature interaction for particles in
the Kerr background while in circular orbit \cite{Mashhoon2}.
From previous work on chaotic dynamics \cite{Suzuki1,Suzuki2,Hartl1,Hartl2}, there is a strong suggestion
that perturbations in the M{\o}ller radius may give some insight into determining the precise conditions for a transition away from
stable motion.
A straightforward calculation shows that
\be
{s(\varepsilon) \over m(\varepsilon)}
& = & \varepsilon \, {s_0 \over m_0} \lt\{1 + \varepsilon \lt[{1 \over 2} \lt(\bar{s}_1^2 - \bar{m}_1^2\rt) \rt] \rt.
\nn
&  &{} + \varepsilon^2 \lt[{1 \over 2} \lt(\bar{s}_2^2 - \bar{m}_2^2\rt) - {1 \over 4} \, \bar{s}_1^2 \, \bar{m}_1^2
- {1 \over 8} \lt(\bar{s}_1^4 - 3 \, \bar{m}_1^4\rt)\rt]
\nn
&  &{} + \lt. O(\varepsilon^3) \rt\},
\label{s/m}
\ee
where the second- and third-order contributions in $\varepsilon$ due to the ``radiative corrections'' formally
shift the perturbed terms away from $\rho_0 = s_0/m_0$.
However, the precise nature of the shift from $\rho_0$ to $\rho$ requires determining (\ref{s/m})
in terms of a specific background.

\subsection{Solving for Linear Momentum and Spin Angular Momentum Expansion Components}
\label{sec:3.3}

\subsubsection{Local Fermi Co-ordinate Frame}

Having now presented the generalized CMP approximation, the next step is to determine the linear momentum
and spin angular momentum series expansion components for (\ref{P-S-approx}).
This can be accomplished iteratively by solving the first-order perturbations with respect to zeroth-order
quantities, and then computing the higher-order terms in a systematic fashion.
This approach becomes particularly straightforward upon framing the problem in terms of the tetrad formalism
and Fermi normal co-ordinates \cite{Mashhoon2}, the latter of which has the property that the corresponding metric
in the neighbourhood of a freely falling worldline is locally flat.
The leading-order metric deviations in Fermi normal co-ordinates are then proportional
to the projected Riemann curvature tensor in the Fermi frame evaluated on the worldline.

To begin, consider an orthonormal tetrad frame $\lm^\mu{}_{\hat{\al}}$ with the orthogonality condition
\be
\eta_{\hat{\al} \hat{\bt}} & = & g_{\mu \nu} \, \lm^\mu{}_{\hat{\al}} \, \lm^\nu{}_{\hat{\bt}} \,
\label{ortho-tetrad}
\ee
satisfying parallel transport
\be
{D \lm^\mu{}_{\hat{\al}} \over \d \tau} & = & 0 \,
\label{tetrad-parallel-transport}
\ee
with respect to the general space-time co-ordinates $X^\mu$, where $\hat{\al}$ are indices
for the Fermi co-ordinates $X^{\hat{\al}}$ defined in the neighbourhood of the spinning particle on a locally flat tangent space.
Furthermore, the Riemann curvature tensor in the Fermi frame is described by
\be
{}^F{}R_{\hat{\al} \hat{\bt} \hat{\gm} \hat{\dl}} & = & R_{\mu \nu \rho \sg} \,
\lm^\mu{}_{\hat{\al}} \, \lm^\nu{}_{\hat{\bt}} \, \lm^\rho{}_{\hat{\gm}} \, \lm^\sg{}_{\hat{\dl}} \, .
\label{F-Riemann}
\ee
By identifying $\lm^\mu{}_{\hat{0}} = u_{(0)}^\mu$ in the usual fashion and making use of (\ref{ortho-tetrad}) and
the first-order spin condition, it follows naturally that
\begin{subequations}
\label{P0-S0=}
\be
P_{(0)}^\mu & = & \lm^\mu{}_{\hat{\al}} \, P_{(0)}^{\hat{\al}} \ = \ m_0 \, \lm^\mu{}_{\hat{0}} \, ,
\label{P0=}
\nl
S_{(0)}^{\mu \nu} & = & \lm^\mu{}_{\hat{\imath}} \, \lm^\nu{}_{\hat{\jmath}} \, S_{(0)}^{\hat{\imath} \hat{\jmath}} \, ,
\label{S0=}
\ee
\end{subequations}
where $P_{(0)}^{\hat{\al}} =  m_0 \, \dl^{\hat{\al}}{}_{\hat{0}}$ and $S_{(0)}^{\hat{\imath} \hat{\jmath}}$ is a constant-valued
spatial antisymmetric tensor whose components are determined from initial conditions.

\subsubsection{Leading Perturbation of Linear Momentum and Spin Angular Momentum}

Determining the first-order perturbation in $\varepsilon$ for the linear momentum is very straightforward.
Given (\ref{DP-1}) and (\ref{tetrad-parallel-transport}), it is shown that
$D P_{(1)}^\mu / \d \tau = \lm^\mu{}_{\hat{\al}} \lt(\d P_{(1)}^{\hat{\al}} / \d \tau \rt)$, leading to
\be
{\d P_{(1)}^{\hat{\al}} \over \d \tau} & = & -{1 \over 2} \, {}^F{}R^{\hat{\al}}{}_{\hat{0} \hat{\imath} \hat{\jmath}} \,
S_{(0)}^{\hat{\imath} \hat{\jmath}} \, ,
\label{dP/dt}
\ee
which can be integrated immediately with the final result of
\be
P_{(1)}^\mu & = & -{1 \over 2} \, \lm^\mu{}_{\hat{k}}
\int \lt({}^F{}R^{\hat{k}}{}_{\hat{0} \hat{\imath} \hat{\jmath}} \, S_{(0)}^{\hat{\imath} \hat{\jmath}}\rt) \d \tau \, .
\label{P1-general}
\ee
It is interesting to note that, when (\ref{P1-general}) is contracted with $P^{(0)}_\mu$, the first-order mass shift contribution is
identically
\be
\bar{m}_1^2 & = & 0 \,
\label{m1-bar^2=0}
\ee
for a general space-time background, leading to simplified expressions for (\ref{MPD-velocity-explicit}) and (\ref{s/m}).

The corresponding expression for the spin tensor, in contrast to the linear momentum, is somewhat more complicated to determine.
It is important to first note that, from (\ref{DS-j}) for $j = 2$,
\be
{D S_{(1)}^{\mu \nu} \over \d \tau} & = & 0 \, .
\label{DS1/dt=0}
\ee
In terms of the tetrad projection,
\be
S_{(1)}^{\mu \nu} & = & \lm^\mu{}_{\hat{\al}} \, \lm^\nu{}_{\hat{\bt}} \, S_{(1)}^{\hat{\al} \hat{\bt}} \,
\nn
& = & 2 \, \lm^{[\mu}{}_{\hat{0}} \, \lm^{\nu]}{}_{\hat{\jmath}} \, S_{(1)}^{\hat{0} \hat{\jmath}}
+ \lm^\mu{}_{\hat{\imath}} \, \lm^\nu{}_{\hat{\jmath}} \, S_{(1)}^{\hat{\imath} \hat{\jmath}} \, ,
\label{S1-tetrad=}
\ee
where it follows from (\ref{s.p=0}) for $j = 1$ that
\be
S_{(1)}^{\hat{0} \hat{\jmath}} & = & -{1 \over m_0} \, P^{(1)}_{\hat{\imath}} \, S_{(0)}^{\hat{\imath} \hat{\jmath}} \, .
\label{S1-0j=}
\ee
As for the components $S_{(1)}^{\hat{\imath} \hat{\jmath}}$ in  (\ref{S1-tetrad=}), they can be formally determined from using
(\ref{s-bar-sq}) for $j = 1$, such that
\be
S_{(1)}^{\hat{\imath} \hat{\jmath}} & = & {1 \over 4} \, \bar{s}_1^2 \, S_{(0)}^{\hat{\imath} \hat{\jmath}} \, .
\label{S1-ij=}
\ee
This, however, leads to a difficulty, in that (\ref{S1-tetrad=}) is still dependent on a yet {\em undetermined parameter} $\bar{s}_1^2$.
While it is tempting to set $\bar{s}_1^2 = 0$ in analogy with $\bar{m}_1^2 = 0$, this is not well justified considering that
$\bar{s}_1^2$ only needs to be {\em covariantly constant} following (\ref{Ds/dt-Dm/dt=0}), and not necessarily zero.

Fortunately, $\bar{s}_1^2$ can be determined separately by directly solving (\ref{DS1/dt=0}), following a variation of an approach
presented earlier \cite{Singh0}.
Making use of the spin condition constraint equation with (\ref{s.p=0}) for $j = 1$, there exist four equations
\be
A^\mu \, S^{(1)}_{\mu \nu} - B_\nu & = & 0 \, ,
\label{S1-4-equations}
\ee
where
\begin{subequations}
\label{A-B=}
\be
A^\mu & \equiv & P^\mu_{(0)} \, ,
\label{A-P}
\nl
\nn
B_\nu & \equiv & -P^\mu_{(1)} \, S^{(0)}_{\mu \nu} \, .
\label{B-P}
\ee
\end{subequations}
With (\ref{S1-4-equations}), it is possible to algebraically solve for the $S^{(1)}_{0j}$ components in terms of the
purely spatial components $S^{(1)}_{ij}$, such that
\begin{subequations}
\label{S-0j=}
\be
S^{(1)}_{01} & = & {1 \over A^0} \lt[A^2 \, S^{(1)}_{12} - A^3 \, S^{(1)}_{31} + B_1\rt] \, ,
\label{S-01}
\nl
S^{(1)}_{02} & = & {1 \over A^0} \lt[A^3 \, S^{(1)}_{23} - A^1 \, S^{(1)}_{12} + B_2\rt] \, ,
\label{S-02}
\nl
S^{(1)}_{03} & = & {1 \over A^0} \lt[A^1 \, S^{(1)}_{31} - A^2 \, S^{(1)}_{23} + B_3\rt] \, .
\label{S-03}
\ee
\end{subequations}
The remaining three components can then be determined by solving (\ref{DS1/dt=0}) in covariant form
for the spatial components, leading to
\be
{D S^{(1)}_{ij} \over \d \tau} & = &
{\d S^{(1)}_{ij} \over \d \tau} + 2 \, u_{(0)}^\al \, \Gm^\bt{}_{\al[i} \, S^{(1)}_{j]\bt} \ = \ 0 \, .
\label{DS1/dt=0-ij}
\ee
Upon substituting (\ref{S-0j=}) into (\ref{DS1/dt=0-ij}), there now exists a first-order inhomogeneous matrix differential
equation to solve, with components
\begin{subequations}
\label{dS1-matrix=}
\be
{\d S^{(1)}_{12}(\tau) \over \d \tau} + {1 \over 2} \, \al^{ij} \, S^{(1)}_{ij}(\tau) \ = \ \dl_{12}(\tau) \, ,
\label{dS1-12=}
\nl
{\d S^{(1)}_{23}(\tau) \over \d \tau} + {1 \over 2} \, \bt^{ij} \, S^{(1)}_{ij}(\tau) \ = \ \dl_{23}(\tau) \, ,
\label{dS1-23=}
\nl
{\d S^{(1)}_{31}(\tau) \over \d \tau} + {1 \over 2} \, \gm^{ij} \, S^{(1)}_{ij}(\tau) \ = \ \dl_{31}(\tau) \, ,
\label{dS1-31=}
\ee
\end{subequations}
where the antisymmetric spatial tensors $\al^{ij}$, $\bt^{ij}$, and $\gm^{ij}$ may each be $\tau$-dependent,
depending on the choice of metric, and the $\dl_{ij}(\tau)$ are due to (\ref{B-P}).

\subsubsection{Higher-Order Perturbation Terms}

Proceeding beyond the leading-order perturbations of $P^\mu$ and $S^{\al \bt}$ is very straightforward.
For the linear momentum, the second-order expression is
\begin{widetext}
\be
P_{(2)}^\mu & = & -{1 \over 2} \, \lm^\mu{}_{\hat{\al}}
\int \lt({1 \over m_0} \, {}^F{}R^{\hat{\al}}{}_{\hat{\bt} \hat{k} \hat{l}} \, P_{(1)}^{\hat{\bt}} \, S_{(0)}^{\hat{k} \hat{l}}
+ {}^F{}R^{\hat{\al}}{}_{\hat{0} \hat{\gm} \hat{\bt}} \, S_{(1)}^{\hat{\gm} \hat{\bt}}\rt) \d \tau
\nn
& \approx & -{1 \over 2} \, \lm^\mu{}_{\hat{\al}}
\int \lt({1 \over m_0} \, {}^F{}R^{\hat{\al}}{}_{\hat{\bt} \hat{k} \hat{l}} \, P_{(1)}^{\hat{\bt}}
+ {1 \over 4} \, \lt\langle \bar{s}_1^2 \rt\rangle \, {}^F{}R^{\hat{\al}}{}_{\hat{0} \hat{k} \hat{l}}
- {2 \over m_0} \, {}^F{}R^{\hat{\al}}{}_{\hat{0} \hat{0} \hat{l}} \, P^{(1)}_{\hat{k}} \rt) S_{(0)}^{\hat{k} \hat{l}} \, \d \tau
\, ,
\label{P2-general}
\ee
\end{widetext}
where
\be
\lt\langle \bar{s}_j^2 \rt\rangle & = & {1 \over T} \, \int_0^T \bar{s}_j^2(\tau) \, \d \tau
\label{<sj^2>}
\ee
is the time-averaged jth-order correction to the squared spin magnitude.
To determine the second-order spin tensor, it is first shown from (\ref{DS-j}) for $j = 3$ that
\be
{D S_{(2)}^{\mu \nu} \over \d \tau} & = & {1 \over m_0^3} \, P_{(0)}^{[\mu} \, S_{(0)}^{\nu]\sg} \, R_{\sg \gm \al \bt} \, P_{(0)}^\gm \,
S_{(0)}^{\al \bt} \, .
\label{DS2/dt=0}
\ee
When expressed in terms of the orthonormal tetrad, it follows that (\ref{DS2/dt=0}) can be solved easily to obtain
\be
S_{(2)}^{\mu \nu} & = & {1 \over m_0} \, \lm^{[\mu}{}_{\hat{0}} \, \lm^{\nu]}{}_{\hat{\imath}}
\int S_{(0)}^{\hat{\imath} \hat{\jmath}} \, {}^F{}R_{\hat{\jmath} \hat{0} \hat{k} \hat{l}} \, S_{(0)}^{\hat{k} \hat{l}} \, \d \tau \, .
\label{S2-general}
\ee


\section{Circular Motion Near the Event Horizon of a Kerr Black Hole}
\label{sec:4}

With the generalized CMP approximation of the MPD equations established, it can now be applied
to the concrete example of a spinning particle in circular motion around a Kerr black hole
in the equatorial plane near its event horizon.
It would be interesting to identify the general interplay between black hole spin and the dynamical
response from the spinning particle.
Since it is known that an orbiting particle whose spin co-rotates with the black hole spin leads
to a repulsive force, while one of opposite spin to the black hole leads to an attractive force \cite{Chicone},
it is particularly useful to determine the black hole's spin dependence on the kinematics and dynamics of the
particle's orbit due to contributions beyond the CMP approximation considered earlier \cite{Mashhoon2}.

To begin, consider a Kerr background in standard Boyer-Lindquist co-ordinates $X^\mu = (t, r, \th, \ph)$,
described in terms of a black hole mass $M$ and specific spin angular momentum $a = J/M$.
Then, for constant radius $r$, the solution to the geodesic equation for circular motion is \cite{Mashhoon2}
\begin{subequations}
\label{t-ph}
\be
t & = & {1 \over N} \lt(1 + a \, \OmK\rt) \tau \, ,
\label{t=}
\nl
\qquad \ph & = & {\OmK \, \tau \over N \, \sin \th}  \, ,
\label{ph=}
\ee
\end{subequations}
where $\th = \pi/2$ on the equatorial plane,
\be
\OmK & = & \sqrt{M \over r^3}
\label{Om-Kepler}
\ee
is the Keplerian frequency of the orbit and
\be
N & = & \sqrt{1 - {3 M \over r} + 2 a \, \OmK}
\label{N=}
\ee
is a normalization constant.
At $\tau = 0$, the boundary conditions are chosen such that $t = \ph = 0$.
Given that $\lm^\mu{}_{\ze} = \d X^\mu/\d \tau$ is the particle's four-velocity vector, and that the orthonormal
tetrad frame satisfies $D \lm^\mu{}_{\Cal}/ \d \tau = 0$, it is possible to determine
unit gyro axes $\lm^\mu{}_{\Ci}$ to describe the particle's local spatial frame.
It is straightforward to show that \cite{Mashhoon2,Chicone1}
\begin{subequations}
\label{tetrad}
\be
\lm^\mu{}_{\hat{0}} & = & \lt({1 + a \, \OmK \over N}, 0, 0, {\OmK \over N \, \sin \th} \rt),
\label{tetrad-0}
\nl
\lm^\mu{}_{\hat{1}} & = & \lt(-{L \over r \, A} \, \sin \lt(\OmK \, \tau\rt), \, A \, \cos \lt(\OmK \, \tau\rt), \, 0, \, \rt.
\nn
&  &{} - \lt. {E \over r \, A \, \sin \th} \, \sin \lt(\OmK \, \tau\rt) \rt),
\label{tetrad-1}
\nl
\lm^\mu{}_{\hat{2}} & = & \lt(0, 0, {1 \over r}, 0\rt),
\label{tetrad-2}
\nl
\lm^\mu{}_{\hat{3}} & = & \lt({L \over r \, A} \, \cos \lt(\OmK \, \tau\rt), \, A \, \sin \lt(\OmK \, \tau\rt), \, 0, \, \rt.
\nn
&  &{} \lt. {E \over r \, A \, \sin \th} \, \cos \lt(\OmK \, \tau\rt) \rt),
\label{tetrad-3}
\ee
\end{subequations}
where
\begin{subequations}
\label{E-L-A}
\be
E & = & {1 \over N} \lt(1 - {2M \over r} + a \, \OmK\rt) \, ,
\label{E}
\nl
L & = & {r^2 \OmK \over N} \lt(1 - 2a \, \OmK + {a^2 \over r^2}\rt) \, ,
\label{L}
\nl
A & = & \sqrt{1 - {2M \over r} + {a^2 \over r^2}} \, ,
\label{A}
\ee
\end{subequations}
for the circular orbit's energy $E$ and orbital angular momentum $L$.
It is useful to describe $a$ by the dimensionless parameter $\al = a/r$, such that $-1/4 \leq \al \leq 1$ to incorporate black hole spin
($-M \leq a \leq M$) that is both co-rotating and counter-rotating with respect to the orbital direction,
where $r_{0+} = M$ for $\al = 1$ and $r_{0-} = 4 \, M$ for $\al = -1/4$, each corresponding to the innermost
(photon) radius allowed \cite{Chandra}.
All dimensional quantities are then described with respect to $M$, such that
\be
r & = & 9 \, M \lt[\al + \sqrt{\al^2 + 3 \lt(1 - N^2\rt)}\rt]^{-2} \,
\label{r}
\ee
is the orbital radius near the event horizon and $N \gtrsim 0$ denotes the separation away from the innermost circular orbit.
The parameters (\ref{E-L-A}) are now expressible in terms of
\begin{subequations}
\label{E0-L0-Delta}
\be
E_0 & = & 1 - 2 \, (r^2 \, \OmK^2) + \al \, (r \, \OmK)  \ = \ N \, E \, ,
\label{E0}
\nl
L_0 & = & (r \, \OmK) \lt[1 - 2 \, \al \, (r \, \OmK) + \al^2 \rt]
\nn
& = & N \, {L \over r} \, ,
\label{L0}
\ee
\end{subequations}
where $\Dl = r^2 \lt[1 - 2 \, (r^2 \, \OmK^2) + \al^2\rt] = r^2 \, A^2$ is the known function defined in the Kerr metric \cite{Chandra}.

The exact expressions for ${}^F{}R_{\Cmu \Cnu \Cal \Cbt}$ are listed in Appendix~\ref{appendix:riemann-frame-components}.
For the special case of $N \rightarrow 0$ and $\th = \pi/2$ considered in this paper, the dominant nonzero contributions of the
Riemann curvature tensor ${}^F{}R_{\Cmu \Cnu \Cal \Cbt}$ in the Fermi frame are
\begin{widetext}
\begin{subequations}
\label{Riemann-Fermi}
\be
{}^F{}R_{\ze \on \ze \on} & \approx & -{\OmK^2 \over N^2} \,
\lt[2 \, A^2 + r^2 \, \OmK^2 - \al \lt(2 \, r \, \OmK - \al\rt) \rt] \, \cos^2 \lt(\OmK \, \tau\rt)
\ = \ - {}^F{}R_{\tw \tr \tw \tr} \, ,
\nl
{}^F{}R_{\ze \on \ze \tr} & \approx & -{\OmK^2 \over N^2} \,
\lt[2 \, A^2 + r^2 \, \OmK^2 - \al \lt(2 \, r \, \OmK - \al\rt) \rt] \, \sin \lt(\OmK \, \tau \rt) \, \cos \lt(\OmK \, \tau\rt)
\ = \ - {}^F{}R_{\on \tw \tw \tr} \, ,
\nl
{}^F{}R_{\ze \on \on \tr} & \approx & {3 \, \OmK^2 \, A \over N^2} \, \lt(r \, \OmK - \al\rt) \, \cos \lt(\OmK \, \tau\rt)
\ = \ - {}^F{}R_{\ze \tw \tw \tr} \, ,
\nl
{}^F{}R_{\ze \tw \ze \tw} & \approx & {\OmK^2 \over N^2} \, \lt[1 - 2 \, \al \lt(2 \, r \, \OmK - \al\rt) \rt]
\ = \ - {}^F{}R_{\on \tr \on \tr} \, ,
\nl
{}^F{}R_{\ze \tw \on \tw} & \approx & - {3 \, \OmK^2 \, A \over N^2} \, \lt(r \, \OmK - \al\rt) \, \sin \lt(\OmK \, \tau\rt)
\ = \ - {}^F{}R_{\ze \tr \on \tr} \, ,
\nl
{}^F{}R_{\ze \tr \ze \tr} & \approx & -{\OmK^2 \over N^2} \,
\lt[2 \, A^2 + r^2 \, \OmK^2 - \al \lt(2 \, r \, \OmK - \al\rt) \rt]  \, \sin^2 \lt(\OmK \, \tau\rt)
\ = \ - {}^F{}R_{\on \tw \on \tw} \, .
\ee
\end{subequations}
\end{widetext}

\subsection{First-Order Perturbations in $\varepsilon$}
\label{sec:4.1}

With all the ground work for the application of the generalized CMP approximation now complete,
it is possible to begin computing the linear and higher-order perturbations of the linear momentum
and spin tensor for a spinning point particle in circular orbit.
In this context, the expansion parameter $\varepsilon$ is associated with the unperturbed M{\o}ller radius $\rho_0 = s_0/m_0$
in unit of $r$, such that $s_0/(m_0 \, r) \ll 1$.

To begin, recall from (\ref{P0=}) that (\ref{tetrad-0}) that the unperturbed four-momentum components are
\begin{subequations}
\label{P0-components}
\be
P_{(0)}^0 & = & {m_0 \over N} \lt(1 + \al \, r \, \OmK\rt) \, , \qquad
\label{P0-0}
\nl
P_{(0)}^1 & = & 0 \, , \qquad
\label{P0-1}
\nl
P_{(0)}^2 & = & 0 \, , \qquad
\label{P0-2}
\nl
P_{(0)}^3 & = & {m_0 \over N} \, \OmK \, .
\label{P0-3}
\ee
\end{subequations}
%
%
Obtaining the first-order perturbation (CMP approximation) is very straightforward.
Given (\ref{S0=}) and noting that the spinning particle is initially located on the $x$-axis of the Cartesian frame,
the initial spin orientation $(\hat{\th}, \hat{\ph})$ for $S_{(0)}^{\mu \nu}$
coincides with the standard definition for the spherical co-ordinates $(\th, \ph)$ with respect to the Cartesian frame's $z$-axis.
Therefore, it follows that the projected spin tensor components are \cite{Mashhoon2}
\begin{subequations}
\label{S0-proj}
\be
S_{(0)}^{\tw \tr} & = & s_0 \, \sin \hat{\th} \, \cos \hat{\ph} \, ,
\label{S0-23-proj}
\nl
S_{(0)}^{\tr \on} & = & -s_0 \, \cos \hat{\th} \, ,
\label{S0-31-proj}
\nl
S_{(0)}^{\on \tw} & = & s_0 \, \sin \hat{\th} \, \sin \hat{\ph} \, ,
\label{S0-12-proj}
\ee
\end{subequations}
leading to
\begin{widetext}
\begin{subequations}
\label{S0}
\be
S_{(0)}^{01}(\tau) & = & -{m_0 \, r \, L_0 \over N} \lt(s_0 \over m_0 \, r \rt)
\, \cos \hat{\th} \, ,
\label{S0-01}
\nl
S_{(0)}^{02}(\tau) & = & -{1 \over 2} \, {m_0 \, L_0 \over N \, A} \lt(s_0 \over m_0 \, r \rt)
\, \lt[\sin (\OmK \, \tau + \hat{\th} - \hat{\ph}) - \sin (\OmK \, \tau - \hat{\th} - \hat{\ph})\rt] \, ,
\label{S0-02}
\nl
S_{(0)}^{03}(\tau) & = & 0 \, ,
\label{S0-03}
\nl
S_{(0)}^{12}(\tau) & = & {1 \over 2} \, m_0 \, A \, \lt(s_0 \over m_0 \, r \rt)
\lt[\cos (\OmK \, \tau + \hat{\th} - \hat{\ph}) - \cos (\OmK \, \tau - \hat{\th} - \hat{\ph})\rt] \, ,
\label{S0-12}
\nl
S_{(0)}^{23}(\tau) & = & {1 \over 2} \, {m_0 \, E_0 \over r \, N \, A} \lt(s_0 \over m_0 \, r \rt)
\lt[\sin (\OmK \, \tau + \hat{\th} - \hat{\ph}) - \sin (\OmK \, \tau - \hat{\th} - \hat{\ph})\rt] \, ,
\label{S0-23}
\nl
S_{(0)}^{31}(\tau) & = & -{m_0 \, E_0 \over N} \lt(s_0 \over m_0 \, r \rt) \, \cos \hat{\th} \, .
\label{S0-31}
\ee
\end{subequations}
\end{widetext}
It is interesting to note the appearance of a beat structure in the sinusoidal functions of (\ref{S0}), due to
the initial spin orientation angles.
As for the first-order perturbation for the linear momentum, this follows naturally from (\ref{P1-general}), resulting in
\begin{widetext}
\begin{subequations}
\label{P1-components}
\be
P_{(1)}^0(\tau) & = & {3 \over 2} \, {m_0 \, L_0 \over N^3} \, \lt(s_0 \over m_0 \, r \rt) \,
(r \, \OmK) \lt(r \, \OmK - \al\rt)
\lt[\cos (\OmK \, \tau + \hat{\th}) + \cos (\OmK \, \tau - \hat{\th}) - 2 \, \cos \hat{\th} \rt] \, ,
\label{P1-0}
\nl
P_{(1)}^1(\tau) & = & {3 \over 2} \, {m_0 \, A^2 \over N^2} \, \lt(s_0 \over m_0 \, r \rt) \,
(r \, \OmK) \lt(r \, \OmK - \al\rt) \lt[\sin (\OmK \, \tau + \hat{\th}) + \sin (\OmK \, \tau - \hat{\th}) \rt] \, ,
\label{P1-1}
\nl
P_{(1)}^2(\tau) & = & {3 \over 2} \, {m_0 \, A \over r \, N^2} \, \lt(s_0 \over m_0 \, r \rt) \,
(r \, \OmK) \lt(r \, \OmK - \al\rt)
\lt[\cos (\OmK \, \tau + \hat{\th} - \hat{\ph}) - \cos (\OmK \, \tau - \hat{\th} - \hat{\ph}) \rt.
\nn
&  &{} + \lt. \cos(\hat{\th} + \hat{\ph}) - \cos(\hat{\th} - \hat{\ph}) \rt] \, ,
\label{P1-2}
\nl
P_{(1)}^3(\tau) & = & {3 \over 2} \, {m_0 \, E_0 \over r \,  N^3} \, \lt(s_0 \over m_0 \, r \rt) \,
(r \, \OmK) \lt(r \, \OmK - \al\rt) \lt[\cos (\OmK \, \tau + \hat{\th}) + \cos (\OmK \, \tau - \hat{\th}) - 2 \, \cos \hat{\th} \rt] \, ,
\label{P1-3}
\ee
\end{subequations}
\end{widetext}
which also exhibits a beat structure similar to what is found in (\ref{S0}).
It is also interesting to note the relationship between the azimuthal and time component of $P_{(1)}^\mu$ in the form
\be
{P_{(1)}^3(\tau) \over P_{(1)}^0(\tau)}  & = & {E \over L} \, ,
\label{P1-3-P1-0}
\ee
in agreement with the same computation performed earlier \cite{Mashhoon2}.
At this point, it is important to understand the conditions for the collapse of
(\ref{P1-components}) when $\lt(r \, \OmK - \al\rt) = 0$.
This condition can appear when
\be
r_{\rm c} & = & \lt(a \over M\rt)^2 M \, ,
\label{r-c1}
\ee
which is theoretically possible to reach when $a = M$ \cite{Chandra}.
However, it seems unlikely that such a possibility would arise in a realistic astrophysical context.

\subsection{Higher-Order Perturbations in $\varepsilon$}
\label{sec:4.2}

Evaluation of the second-order perturbation quantities is also straightforward, though rather involved.
There are, however, relatively compact expressions for higher-order perturbation quantities in $\varepsilon$ that
are required to evaluate the perturbed M{\o}ller radius (\ref{s/m}), namely the ``radiative corrections'' to the squared mass
and spin magnitudes (\ref{m-bar-sq}) and (\ref{s-bar-sq}).
The first computation of interest is $\bar{s}_1^2$, which is third-order in $\varepsilon$ according to (\ref{s-sq}).
This is achieved by solving the first-order inhomogeneous matrix differential equation (\ref{dS1-matrix=}) for $S^{(1)}_{ij}(\tau)$,
followed by the algebraic equation (\ref{S-0j=}) for the remaining components $S^{(1)}_{0j}(\tau)$, which yields $\bar{s}_1^2$
via (\ref{s-bar-sq}).
Before proceeding, it is useful to first introduce a convenient notation for beat functions taking the form
\begin{subequations}
\label{Q-cs}
\be
Q_{\rm c}^\pm (n_1 \, \OmK \, \tau \, ,  n_2 \, \hat{\th} \, , \, n_3 \, \hat{\ph}) & \equiv &
\cos (n_1 \OmK \, \tau + n_2 \, \hat{\th} - n_3 \, \hat{\ph})
\nn
& \pm & \cos (n_1 \OmK \, \tau - n_2 \, \hat{\th} - n_3 \, \hat{\ph}) \, ,
\nn
\label{Q-c}
\nl
Q_{\rm s}^\pm (n_1 \, \OmK \, \tau \, ,  n_2 \, \hat{\th} \, , \, n_3 \, \hat{\ph}) & \equiv &
\sin (n_1 \OmK \, \tau + n_2 \, \hat{\th} - n_3 \, \hat{\ph})
\nn
& \pm & \sin (n_1 \OmK \, \tau - n_2 \, \hat{\th} - n_3 \, \hat{\ph}) \, ,
\nn
\label{Q-s}
\ee
\end{subequations}
which appear frequently in subsequent expressions throughout this paper.
For the Kerr metric, it is shown that the nonzero $\al^{ij}$, $\bt^{ij}$, and $\gm^{ij}$ for (\ref{dS1-matrix=}) are
\begin{subequations}
\label{al-bt-nonzero}
\be
\al^{23} & = & {N \over r \, A^2} \, {\OmK \over \lt(1 + \al \, r \, \OmK\rt)} \, ,
\label{al-23}
\nl
\bt^{12} & = & -{A^2 \over N} \, (r \, \OmK) \lt(1 + \al \, r \, \OmK\rt) \, ,
\label{bt-12}
\ee
\end{subequations}
while
\begin{subequations}
\label{dl}
\be
\dl_{12}(\tau) & = & -{3 \over 4} \, {m_0 \, r  \, L_0 \over N^2 \, A} \,
\lt(s_0 \over m_0 \, r \rt)^2  \, {(r^2 \, \OmK^2) \lt(r \, \OmK - \al\rt) \over  \lt(1 + \al \, r \, \OmK\rt)} \,
\nn
&  &{} \times Q_{\rm s}^- (\OmK \, \tau \, ,  2 \, \hat{\th} \, , \hat{\ph}) \, ,
\label{dl-12}
\nl
\dl_{23}(\tau) & = & 0 \, ,
\label{dl-23}
\nl
\dl_{31}(\tau) & = & {3 \over 8} \, {m_0 \, r \, L_0 \over N^3} \,
\lt(s_0 \over m_0 \, r \rt)^2 \, (r^2 \, \OmK^2) \lt(r \, \OmK - \al\rt)
\nn
&  &{} \times
\lt[Q_{\rm s}^+ (2 \, \OmK \, \tau \, ,  2 \, \hat{\th} \, , 2 \, \hat{\ph})
- \sin (2 \, \OmK \, \tau - 2 \, \hat{\ph}) \rt] \, .
\nn
\label{dl-31}
\ee
\end{subequations}
Then the solutions to (\ref{S-0j=}) and (\ref{dS1-matrix=}) are
\begin{widetext}
\begin{subequations}
\label{S1-cov}
\be
S^{(1)}_{01}(\tau) & = & {3 \over 16} \, m_0 \, r \, \lt(s_0 \over m_0 \, r \rt)^2 \,
{L_0 \over N^3} \, {(r^2 \, \OmK^2) \lt(r \, \OmK - \al\rt) \over  \lt(1 + \al \, r \, \OmK\rt)}
\nn
&  &{} \times \lt\{Q_{\rm c}^+ (2 \, \OmK \, \tau \, ,  2 \, \hat{\th} \, , 2 \, \hat{\ph})
- Q_{\rm c}^+ (0 \, ,  2 \, \hat{\th} \, , 2 \, \hat{\ph})
- 2 \lt[ \cos (2 \, \OmK \, \tau - 2 \, \hat{\ph}) - \cos (2 \, \hat{\ph}) \rt] \rt\} \, ,
\label{S1-01-cov}
\nl
S^{(1)}_{02}(\tau) & = & {3 \over 16} \, m_0 \, r^2 \, \lt(s_0 \over m_0 \, r \rt)^2 \,
{L_0 \, A \over N^3} \, {(r^2 \, \OmK^2) \lt(r \, \OmK - \al\rt) \over  \lt(1 + \al \, r \, \OmK\rt)}
\nn
&  &{} \times \lt\{\lt(2 \,  \, \OmK \, \tau \rt) Q_{\rm c}^- (\OmK \, \tau \, ,  2 \, \hat{\th} \, , \hat{\ph})
- \lt[Q_{\rm s}^- (\OmK \, \tau \, ,  2 \, \hat{\th} \, , \hat{\ph}) - Q_{\rm s}^- (0 \, ,  2 \, \hat{\th} \, , \hat{\ph}) \rt] \rt\} \, ,
\label{S1-02-cov}
\nl
S^{(1)}_{03}(\tau) & = & -{3 \over 8} \, m_0 \, r^2 \, \lt(s_0 \over m_0 \, r \rt)^2 \,
{A^2 \over N^2} \, (r \, \OmK) \lt(r \, \OmK - \al\rt)
\nn
&  &{} \times \lt\{Q_{\rm s}^+ (2 \, \OmK \, \tau \, ,  2 \, \hat{\th} \, , 2 \, \hat{\ph})
- Q_{\rm s}^+ (0 \, ,  2 \, \hat{\th} \, , 2 \, \hat{\ph})
- 2 \lt[ \sin (2 \, \OmK \, \tau - 2 \, \hat{\ph}) + \sin (2 \, \hat{\ph}) \rt] \rt\} \, ,
\label{S1-03-cov}
\nl
S^{(1)}_{12}(\tau) & = & -{3 \over 16} \, m_0 \, r^2 \, \lt(s_0 \over m_0 \, r \rt)^2 \,
{L_0 \over N^2 \, A} \, {(r \, \OmK) \lt(r \, \OmK - \al\rt) \over  \lt(1 + \al \, r \, \OmK\rt)}
\nn
&  &{} \times \lt\{\lt(2 \,  \, \OmK \, \tau \rt) Q_{\rm s}^- (\OmK \, \tau \, ,  2 \, \hat{\th} \, , \hat{\ph})
- \lt[Q_{\rm c}^- (\OmK \, \tau \, ,  2 \, \hat{\th} \, , \hat{\ph}) - Q_{\rm c}^- (0 \, ,  2 \, \hat{\th} \, , \hat{\ph}) \rt] \rt\} \, ,
\label{S1-12-cov}
\nl
S^{(1)}_{23}(\tau) & = & {3 \over 16} \, m_0 \, r^3 \, \lt(s_0 \over m_0 \, r \rt)^2 \,
{L_0 \, A \over N^3} \, (r \, \OmK) \lt(r \, \OmK - \al\rt)
\nn
&  &{} \times \lt\{\lt(2 \,  \, \OmK \, \tau \rt) Q_{\rm c}^- (\OmK \, \tau \, ,  2 \, \hat{\th} \, , \hat{\ph})
- \lt[Q_{\rm s}^- (\OmK \, \tau \, ,  2 \, \hat{\th} \, , \hat{\ph}) - Q_{\rm s}^- (0 \, ,  2 \, \hat{\th} \, , \hat{\ph}) \rt] \rt\} \, ,
\label{S1-23-cov}
\nl
S^{(1)}_{31}(\tau) & = & -{3 \over 16} \,  m_0 \, r^2 \, \lt(s_0 \over m_0 \, r \rt)^2 \,
{L_0 \over N^3} \, (r \, \OmK) \lt(r \, \OmK - \al\rt)
\nn
&  &{} \times \lt\{Q_{\rm c}^+ (2 \, \OmK \, \tau \, ,  2 \, \hat{\th} \, , 2 \, \hat{\ph})
- Q_{\rm c}^+ (0 \, ,  2 \, \hat{\th} \, , 2 \, \hat{\ph})
- 2 \lt[ \cos (2 \, \OmK \, \tau - 2 \, \hat{\ph}) - \cos (2 \, \hat{\ph}) \rt] \rt\} \, .
\label{S1-31-cov}
\ee
\end{subequations}
\end{widetext}
When combined with (\ref{S0}), it follows that the first-order shift in the squared spin magnitude is
\be
\bar{s}_1^2(\tau) & = & {1 \over N^2} \, \lt(s_0 \over m_0 \, r \rt) \tilde{s}_1^2(\tau) \, , 
\label{s1-bar^2=}
\ee
where
\begin{widetext}
\be
\tilde{s}_1^2(\tau) & = &
{3 \over 16} \, {L_0 \lt(r \, \OmK\rt) \lt(r \, \OmK - \al\rt) \over \lt(1 + \al \, r \, \OmK\rt)}
\lt\{\lt(2 \, \OmK \, \tau\rt) \lt[Q_{\rm s}^+ (2 \, \OmK \, \tau \, ,  3 \, \hat{\th} \, , 2 \, \hat{\ph})
- Q_{\rm s}^+ (2 \, \OmK \, \tau \, , \hat{\th} \, , 2 \, \hat{\ph}) \rt] \rt.
\nn
&  &{} + 3 \lt[Q_{\rm c}^+ (2 \, \OmK \, \tau \, ,  3 \, \hat{\th} \, , 2 \, \hat{\ph})
- Q_{\rm c}^+ (2 \, \OmK \, \tau \, , \hat{\th} \, , 2 \, \hat{\ph}) \rt]
- Q_{\rm c}^+ (\OmK \, \tau \, ,  3 \, \hat{\th} \, , 2 \, \hat{\ph})
\nn
&  &{} + Q_{\rm c}^+ (\OmK \, \tau \, ,  \hat{\th} \, , 2 \, \hat{\ph})
- Q_{\rm c}^+ (\OmK \, \tau \, , 3 \, \hat{\th} \, , 0) + Q_{\rm c}^+ (\OmK \, \tau \, , \hat{\th} \, , 0)
\nn
&  &{} - \lt. 2 \lt[Q_{\rm c}^+ (0 \, ,  3 \, \hat{\th} \, , 2 \, \hat{\ph}) - Q_{\rm c}^+ (0 \, , \hat{\th} \, , 2 \, \hat{\ph}) \rt]
+ 2 \lt[\cos (3 \, \hat{\th}) - \cos \hat{\th} \rt] \rt\} \, .
\label{s1-tilde^2=}
\ee
\end{widetext}
The time-averaged expression for (\ref{s1-bar^2=}) over a cycle defined by the Keplerian frequency is
\be
\lt\langle \bar{s}_1^2 \rt\rangle & = & {1 \over N^2} \, \lt(s_0 \over m_0 \, r \rt) \lt\langle \tilde{s}_1^2\rt\rangle \, ,
\label{s1-bar^2-time-avg=}
\ee
where
\be
\lt\langle \tilde{s}_1^2 \rt\rangle & \equiv & {\OmK \over 2 \, \pi} \int_0^{2 \pi/\OmK} \tilde{s}_1^2(\tau) \, \d \tau
\nn
& = & {3 \over 2} \, {L_0 \lt(r \, \OmK\rt) \lt(r \, \OmK - \al\rt) \over \lt(1 + \al \, r \, \OmK\rt)} \,
\sin^2 \hat{\th} \, \cos \hat{\th}
\nn
&  &{} \times \lt[3 \, \cos (2 \, \hat{\th}) - 1 \rt] \, .
\label{s1-tilde^2-time-avg=}
\ee

It is important to note that both (\ref{s1-tilde^2=}) and (\ref{s1-tilde^2-time-avg=}) are well-behaved for the full range
of $\hat{\th}$ and $\hat{\ph}$.
However, there exists the possibility for singularities to appear in the limit as $\lt(1 + \al \, r \, \OmK\rt) \rightarrow 0$.
This occurs when
\be
r_{\rm c} & \rightarrow & \lt(|a| \over M\rt)^{2/3} M \, . \qquad a < 0 \,
\label{r-c2}
\ee
For the allowed radii permitted in the Kerr metric for photon orbits \cite{Chandra}, it is clear that
$r > r_{\rm c}$ for $a < 0$, so the relevant expressions considered here can never become singular.

It is more straightforward to solve for the remaining contributions to the perturbed M{\o}ller radius, $\bar{m}_2^2$ and $\bar{s}_2^2$,
which are fourth-order perturbations in $\varepsilon$ according to (\ref{m-sq}) and (\ref{s-sq}).
The first nonzero ``radiative correction'' to the squared mass magnitude (\ref{m-sq}) is
\be
\bar{m}_2^2(\tau) & = & {9 \over 8} \, {A^2 \over N^4} \, \lt(s_0 \over m_0 \, r \rt)^2 (r^2 \, \OmK^2) \lt(r \, \OmK - \al\rt)^2
\nn
&  &{} \times \lt\{Q_{\rm c}^+ (2 \, \OmK \, \tau \, , 2 \, \hat{\th} \, , 2 \, \hat{\ph}) - Q_{\rm c}^+ (0 \, , 2 \, \hat{\th} \, , 2 \, \hat{\ph}) \rt.
\nn
&  &{} - \lt. 2 \lt[ \cos (2 \, \OmK \, \tau - 2 \, \hat{\ph}) - \cos (2 \, \hat{\ph})\rt] \rt\} \, ,
\label{m2-bar^2=}
\ee
according to (\ref{m-bar-sq}), where its time-averaged expression following the definition given in (\ref{s1-tilde^2-time-avg=}) is
\be
\lt\langle \bar{m}_2^2 \rt\rangle & = &
{9 \over 2} \, {A^2 \over N^4} \, \lt(s_0 \over m_0 \, r \rt)^2 (r^2 \, \OmK^2) \lt(r \, \OmK - \al\rt)^2
\, \sin^2 \hat{\th} \,
\nn
&  &{} \times (2 \, \cos^2 \hat{\th} - 1) \, .
\label{m2-bar^2-time-avg=}
\ee
It is clear that the expressions (\ref{m2-bar^2=}) and (\ref{m2-bar^2-time-avg=}) for $\bar{m}_2^2$ have no co-ordinate
singularities due to $\hat{\th}$ and $\hat{\ph}$, nor are there any physical singularities for reasonable choices for $r$.

As for $\bar{s}_2^2$, it follows from evaluating (\ref{s-bar-sq}) for $j = 2$ that
\begin{widetext}
\be
\bar{s}_2^2(\tau) & = & -{3 \, A^2 \over 8 \, N^4} \, \lt(s_0 \over m_0 \, r \rt) \, (r^2 \, \OmK^2) \, \lt(r \, \OmK - \al\rt)
\lt(1 + \al \, r \, \OmK \rt) \lt[Q_{\rm c}^+ (4 \, \OmK \, \tau \, , 3 \, \hat{\th} \, , 2 \, \hat{\ph})
- Q_{\rm c}^+ (4 \, \OmK \, \tau \, , \hat{\th} \, , 2 \, \hat{\ph}) \rt.
\nn
&  &{} - Q_{\rm c}^+ (3 \, \OmK \, \tau \, , 3 \, \hat{\th} \, , 2 \, \hat{\ph})
+ Q_{\rm c}^+ (3 \, \OmK \, \tau \, , \hat{\th} \, , 2 \, \hat{\ph})
+ 3 \, Q_{\rm c}^+ (3 \, \OmK \, \tau \, , 3 \, \hat{\th} \, , 0) + 5 \, Q_{\rm c}^+ (3 \, \OmK \, \tau \, , \hat{\th} \, , 0)
\nn
&  &{} + Q_{\rm c}^+ (\OmK \, \tau \, , 3 \, \hat{\th} \, , -2 \, \hat{\ph})
- 3 \, Q_{\rm c}^+ (\OmK \, \tau \, , 3 \, \hat{\th} \, , 0) - Q_{\rm c}^+ (\OmK \, \tau \, , \hat{\th} \, , -2 \, \hat{\ph})
- 5 \, Q_{\rm c}^+ (\OmK \, \tau \, , \hat{\th} \, , 0)
\nn
&  &{} - \lt. Q_{\rm c}^+ (0 \, , 3 \, \hat{\th} \, , 2 \, \hat{\ph})
+ Q_{\rm c}^+ (0 \, , \hat{\th} \, , 2 \, \hat{\ph}) \rt] \, ,
\label{s2-bar^2=}
\ee
\end{widetext}
whose time-averaged expression is
\be
\lt\langle \bar{s}_2^2 \rt\rangle & = & - {3 \, A^2 \over N^4} \, \lt(s_0 \over m_0 \, r \rt) \, (r^2 \, \OmK^2) \, \lt(r \, \OmK - \al\rt)
\lt(1 + \al \, r \, \OmK \rt)
\nn
&  &{} \times \sin^2 \hat{\th} \, \cos \hat{\th} \lt(2 \, \cos^2 \hat{\ph} - 1\rt) \, .
\label{s2-bar^2-time-avg=}
\ee
Again, (\ref{s2-bar^2=}) and (\ref{s2-bar^2-time-avg=}) for $\bar{s}_2^2$ indicate well-behaved functions for all choices
of $\hat{\th}$ and $\hat{\ph}$, with no possibility of encountering singularities of any kind.
It is interesting to note, however, that $\bar{s}_2^2$ is {\em linear} in $s_0/(m_0 \, r)$, the same order as found
in (\ref{s1-bar^2=}) for $\bar{s}_1^2$, which is somewhat unexpected.
Furthermore, for $N \rightarrow 0$, it appears at first glance that $\bar{s}_2^2$ will dominate over $\bar{s}_1^2$.
Though it is difficult to identify the source for these unusual features, this may be indicative of
another classical analogy to nonrenormalizability in quantum field theory, where higher-order perturbation terms in the
generalized CMP approximation may possibly contribute to {\em all orders} of the expansion for certain quantities.
This is a matter which may require further study in the future.


\section{Numerical Analysis}
\label{sec:5}

At this point, it is useful to consider some numerical analysis of the main expressions for this paper, which
are found in Appendix~\ref{appendix:plots} of this paper.
The purpose behind this procedure is to get a visual sense for how increasing the order of the
perturbation expansion in the generalized CMP approximation influences the predicted physical behaviour
of the spinning particle in the Kerr background.
Since a purely numerical approach to the MPD equations does not allow for the clear identification of
dominant contributions to the particle's orbital motion, this treatment provides an opportunity to
glean some insight as to where a correspondence between the two approaches may occur.

For all plots presented, $r = 6 M$ and $\hat{\th} = \hat{\ph} = \pi/4$.
Particular attention is given to understanding the general stability of the spinning particle's motion while in
circular orbit around a Kerr black hole, especially given the ``radiative corrections'' of the squared mass and spin
magnitudes denoted by (\ref{m-bar-sq}) and (\ref{s-bar-sq}), respectively.

For the purposes of this paper, $\mu \equiv s_0/(m_0 \, r) = 10^{-2}$ and $\mu = 10^{-1}$ are considered throughout,
where $m_0 = 10^{-2} M$.
It so happens that, for the given choices of $r$ and $m_0$, it follows that
\be
s_0 & = & \lt(10^2 \, {r \, \mu \over M} \rt) \, m_0^2 \, .
\label{s0-realistic}
\ee
This implies that a realistic spin of $s_0 \lesssim m_0^2$ for solar mass black holes and neutron stars \cite{Hartl1,Cook}
orbiting supermassive black holes requires that
\be
\mu & \lesssim & 10^{-2} \, {M \over r} \, .
\label{mu-realistic}
\ee
This upper bound given by (\ref{mu-realistic}) indicates that the choices of $\mu = 10^{-2}$ and $\mu = 10^{-1}$
correspond to unrealistically large values \cite{Hartl1,Hartl2} for $s_0$, and consistent with
values chosen in previous work \cite{Suzuki1,Suzuki2} suggesting chaotic behaviour for the MPD equations.
Therefore, any chaotic phenomena reported in this paper occurs under conditions not expected to be realized
in a realistic astrophysical setting.

\begin{figure*}
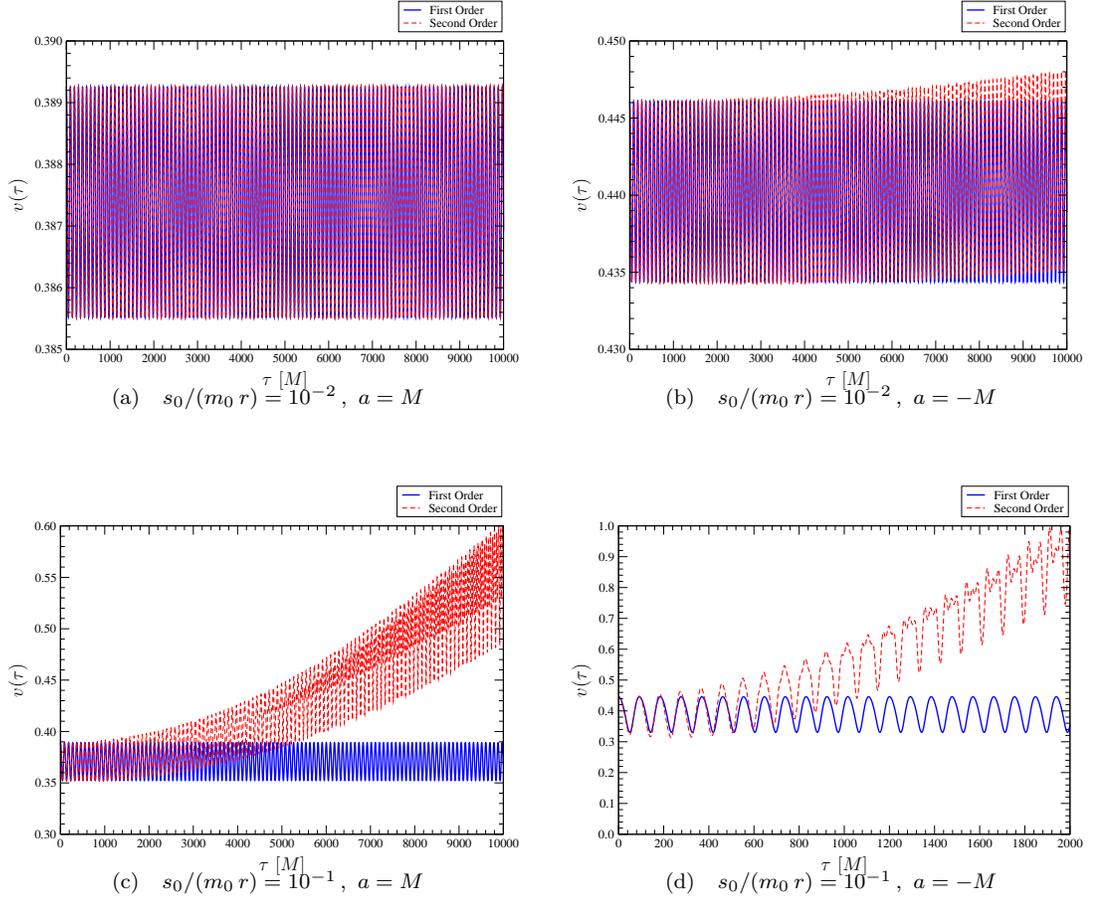

\psfrag{T}[tc][][1.8][0]{\Large $\tau \ [M]$}
\psfrag{v}[bc][][1.8][0]{\Large $v(\tau)$}
\begin{minipage}[t]{0.3 \textwidth}
\centering
\subfigure[\hspace{0.2cm} $s_0/(m_0 \, r) = 10^{-2} \, , \ a = M$]{
\label{fig:v-Kerr-r=6-T=025-P=025-a=+1-mu=1e-2}
\rotatebox{0}{\includegraphics[width = 6.6cm, height = 5.0cm, scale = 1]{1a}}}
\end{minipage}%
\hspace{2.0cm}
\begin{minipage}[t]{0.3 \textwidth}
\centering
\subfigure[\hspace{0.2cm} $s_0/(m_0 \, r) = 10^{-2} \, , \ a = -M$]{
\label{fig:v-Kerr-r=6-T=025-P=025-a=-1-mu=1e-2}
\rotatebox{0}{\includegraphics[width = 6.6cm, height = 5.0cm, scale = 1]{1b}}}
\end{minipage} \\
\vspace{0.8cm}
\begin{minipage}[t]{0.3 \textwidth}
\centering
\subfigure[\hspace{0.2cm} $s_0/(m_0 \, r) = 10^{-1} \, , \ a = M$]{
\label{fig:v-Kerr-r=6-T=025-P=025-a=+1-mu=1e-1}
\rotatebox{0}{\includegraphics[width = 6.6cm, height = 5.0cm, scale = 1]{1c}}}
\end{minipage}%
\hspace{2.0cm}
\begin{minipage}[t]{0.3 \textwidth}
\centering
\subfigure[\hspace{0.2cm} $s_0/(m_0 \, r) = 10^{-1} \, , \ a = -M$]{
\label{fig:v-Kerr-r=6-T=025-P=025-a=-1-mu=1e-1}
\rotatebox{0}{\includegraphics[width = 6.6cm, height = 5.0cm, scale = 1]{1d}}}
\end{minipage}
\caption{\label{fig:v-Kerr-r=6-T=025-P=025} Co-ordinate speed $v(\tau)$ of the spinning particle while in circular orbit around a Kerr black hole,
for $r = 6 M$ and $\hat{\th} = \hat{\ph} = \pi/4$.
It is clear from Fig.~\ref{fig:v-Kerr-r=6-T=025-P=025-a=+1-mu=1e-2} that $v(\tau)$ to second-order contribution in $\varepsilon$
has no significant impact on altering the particle's orbital speed for $s_0/(m_0 \, r) = 10^{-2}$, though
\ref{fig:v-Kerr-r=6-T=025-P=025-a=-1-mu=1e-2} indicates a slight increase over time.
In contrast, Figs.~\ref{fig:v-Kerr-r=6-T=025-P=025-a=+1-mu=1e-1} and \ref{fig:v-Kerr-r=6-T=025-P=025-a=-1-mu=1e-1} show
that an instability occurs as $s_0/(m_0 \, r) = 10^{-1}$ when considering the second-order contribution of $\varepsilon$
in $v(\tau)$.}
\end{figure*}

To begin, consider the magnitude for the particle's co-ordinate speed
\be
v(\varepsilon) & = & \sqrt{g_{ij} \, V^i(\varepsilon) \, V^j(\varepsilon)} \, ,
\label{v-defn}
\ee
where
\be
V^i(\varepsilon) & \equiv & {u^i (\varepsilon) \over u^0 (\varepsilon)} \, ,
\label{v-comp-defn}
\ee
and the $u^\mu(\varepsilon)$ are given by (\ref{MPD-velocity-explicit}).
Since it must be true that $0 \leq v < 1$, it follows that any violation of this range of validity
reflects a breakdown of the formalism's applicability.
An exploration of (\ref{v-defn}) is presented in Figure~\ref{fig:v-Kerr-r=6-T=025-P=025}, to first- and second-order
in $\varepsilon$, for the special cases of co-rotating $(a = M)$ and counter-rotating $(a = -M)$ extreme Kerr
black holes.
It is important to note that while $r = 6 M$ corresponds to stable orbital motion for a spinless particle when $a = M$,
this choice for $r$ only leads to marginally bounded orbits when $a = -M$ \cite{Chandra}.

For Figs.~\ref{fig:v-Kerr-r=6-T=025-P=025-a=+1-mu=1e-2} and \ref{fig:v-Kerr-r=6-T=025-P=025-a=-1-mu=1e-2}, the choice of
$s_0/(m_0 \, r) = 10^{-2}$ shows that the spin-curvature force acting on the particle's motion is almost exclusively due
to the expression to first-order in $\varepsilon$.
In particular, the overall motion is stable throughout the range considered, with a variation on the order of $10^{-3}$
for Fig.~\ref{fig:v-Kerr-r=6-T=025-P=025-a=+1-mu=1e-2} and $10^{-2}$ for Fig.~\ref{fig:v-Kerr-r=6-T=025-P=025-a=-1-mu=1e-2},
where the $O(\varepsilon^2)$ expression only yields a $2 \times 10^{-3}$ increase at the end of the plot
compared to the first-order contribution alone.
However, when $s_0/(m_0 \, r) = 10^{-1}$, the situation changes dramatically for both cases of $a$, as illustrated by
Figs.~\ref{fig:v-Kerr-r=6-T=025-P=025-a=+1-mu=1e-1} and \ref{fig:v-Kerr-r=6-T=025-P=025-a=-1-mu=1e-1}.
This is because the second-order expression in $\varepsilon$ introduces a rapid increase in the co-ordinate speed that
approaches the $v = 1$ upper bound.
It is particularly evident to see this in Fig.~\ref{fig:v-Kerr-r=6-T=025-P=025-a=-1-mu=1e-1}, which formally exceeds
$v(\tau) = 1$ for $\tau > 2000 M$.
Such an outcome for Figs.~\ref{fig:v-Kerr-r=6-T=025-P=025-a=+1-mu=1e-1} and \ref{fig:v-Kerr-r=6-T=025-P=025-a=-1-mu=1e-1}
is consistent with prior numerical analysis on orbital stability \cite{Suzuki1,Hartl1} when considering
large initial spin magnitudes $s_0$.

\begin{figure*}
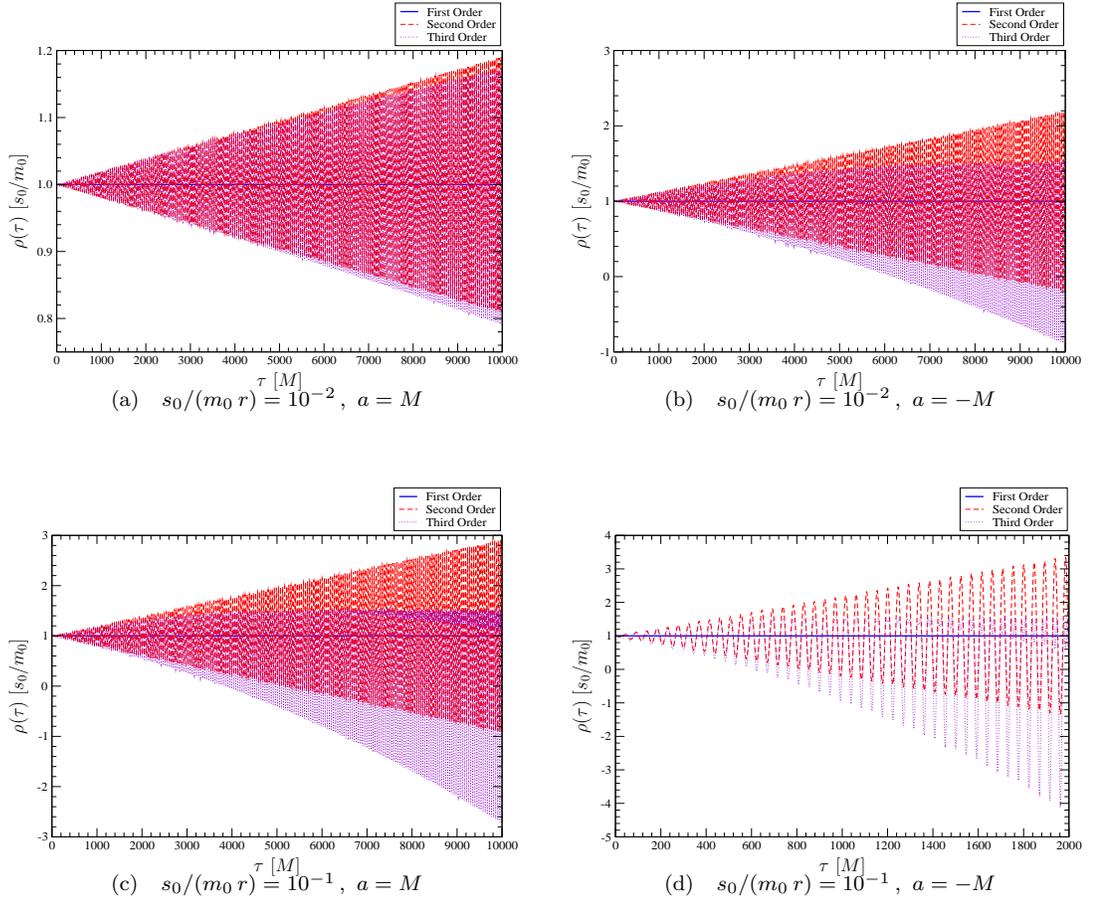

\psfrag{T}[tc][][1.8][0]{\Large $\tau \ [M]$}
\psfrag{s/m}[bc][][1.8][0]{\Large $\rho(\tau) \ [s_0/m_0]$}
\begin{minipage}[t]{0.3 \textwidth}
\centering
\subfigure[\hspace{0.2cm} $s_0/(m_0 \, r) = 10^{-2} \, , \ a = M$]{
\label{fig:moller-Kerr-r=6-T=025-P=025-a=+1-mu=1e-2}
\rotatebox{0}{\includegraphics[width = 6.6cm, height = 5.0cm, scale = 1]{2a}}}
\end{minipage}%
\hspace{2.0cm}
\begin{minipage}[t]{0.3 \textwidth}
\centering
\subfigure[\hspace{0.2cm} $s_0/(m_0 \, r) = 10^{-2} \, , \ a = -M$]{
\label{fig:moller-Kerr-r=6-T=025-P=025-a=-1-mu=1e-2}
\rotatebox{0}{\includegraphics[width = 6.6cm, height = 5.0cm, scale = 1]{2b}}}
\end{minipage} \\
\vspace{0.8cm}
\begin{minipage}[t]{0.3 \textwidth}
\centering
\subfigure[\hspace{0.2cm} $s_0/(m_0 \, r) = 10^{-1} \, , \ a = M$]{
\label{fig:moller-Kerr-r=6-T=025-P=025-a=+1-mu=1e-1}
\rotatebox{0}{\includegraphics[width = 6.6cm, height = 5.0cm, scale = 1]{2c}}}
\end{minipage}%
\hspace{2.0cm}
\begin{minipage}[t]{0.3 \textwidth}
\centering
\subfigure[\hspace{0.2cm} $s_0/(m_0 \, r) = 10^{-1} \, , \ a = -M$]{
\label{fig:moller-Kerr-r=6-T=025-P=025-a=-1-mu=1e-1}
\rotatebox{0}{\includegraphics[width = 6.6cm, height = 5.0cm, scale = 1]{2d}}}
\end{minipage}
\caption{\label{fig:moller-Kerr-r=6-T=025-P=025} M{\o}ller radius $\rho(\tau) = (s/m)(\tau)$ for
$r = 6 M$ and $\hat{\th} = \hat{\ph} = \pi/4$, in units of $s_0/m_0$.
Fig.~\ref{fig:moller-Kerr-r=6-T=025-P=025-a=+1-mu=1e-2} shows that while the higher-order contributions in $\varepsilon$
lead to a slowly increasing amplitude in $\rho$, with a moderate increase found in Fig.~\ref{fig:moller-Kerr-r=6-T=025-P=025-a=-1-mu=1e-2}.
As $s_0/(m_0 \, r) = 10^{-1}$, the amplitude increase becomes more pronounced for
Figs.~\ref{fig:moller-Kerr-r=6-T=025-P=025-a=+1-mu=1e-1} and \ref{fig:moller-Kerr-r=6-T=025-P=025-a=-1-mu=1e-1},
where the second- and third-order contributions in $\varepsilon$ are noticeable.}
\end{figure*}

For immediate comparison, it is useful to now consider the M{\o}ller radius $\rho(\tau) = (s/m)(\tau)$ given by (\ref{s/m})
for the same set of initial conditions.
The purpose of this analysis is to determine whether a correlation exists between the kinematic effects in $v(\tau)$
with the anticipated dynamical contributions due to the spin-curvature interaction in $\rho(\tau)$.
Figure~\ref{fig:moller-Kerr-r=6-T=025-P=025} is a plot of the M{\o}ller radius in units of $s_0/m_0$, up to
$O(\varepsilon^3)$, for $a = M$ and $a = -M$.
Comparison with Figure~\ref{fig:v-Kerr-r=6-T=025-P=025} suggests that such a correlation exists between the two plots.
The expression to third-order in $\varepsilon$, which includes the mass shift contribution $\bar{m}_2^2$ as well as the second-order
spin shift term $\bar{s}_2^2$, has the effect of shifting the range of oscillation downwards within the plots
as compared to the second-order expression alone, which oscillates with growing amplitude about $\rho = 1$.

According to Fig.~\ref{fig:moller-Kerr-r=6-T=025-P=025-a=+1-mu=1e-2} for $s_0/(m_0 \, r) = 10^{-2}$ and $a = M$,
the amplitude for M{\o}ller radius grows very slowly when compared to Fig.~\ref{fig:moller-Kerr-r=6-T=025-P=025-a=-1-mu=1e-2}
for $s_0/(m_0 \, r) = 10^{-2}$ and $a = -M$.
Both the growth of the amplitude and the downward shift of the plots become more pronounced when examining
Figs.~\ref{fig:moller-Kerr-r=6-T=025-P=025-a=+1-mu=1e-1} and \ref{fig:moller-Kerr-r=6-T=025-P=025-a=-1-mu=1e-1} for $a = M$ and $a = -M$,
respectively, when $s_0/(m_0 \, r) = 10^{-1}$.
In particular, the amplitude becomes many times larger than $\rho_0 = s_0/m_0 = 1$ for both the co-rotating and counter-rotating
black hole cases, which suggests that a certain minimum value for $\rho(\tau)$ must occur before
instability of the orbital motion appears.
Though the M{\o}ller radius is not apparently a geometric quantity that must necessarily be positive-valued, it is interesting to
note that the rapid increase in $v(\tau)$ roughly coincides with the condition that $\rho(\tau) < 0$ for each of the plots
in Figure~\ref{fig:moller-Kerr-r=6-T=025-P=025}.
This may be a useful criterion for helping to determine the occurrence of instabilities in the spinning particle's orbital motion.

\begin{figure*}
\psfrag{Th}[cc][][1.8][0]{\Large $\hat{\th}$}
\psfrag{Ph}[cc][][1.8][0]{\Large $\hat{\ph}$}
\psfrag{s/m}[bc][][1.8][90]{\Large $\lt\langle \rho \rt\rangle \ [s_0/m_0]$}
\begin{minipage}[t]{0.3 \textwidth}
\centering
\subfigure[\hspace{0.2cm} $a = M \, , \ O(\varepsilon^2)$]{
\label{fig:avg-Moller-Kerr-a=+1-e2}
\rotatebox{0}{\includegraphics[width = 5.0cm, height = 3.5cm, scale = 1]{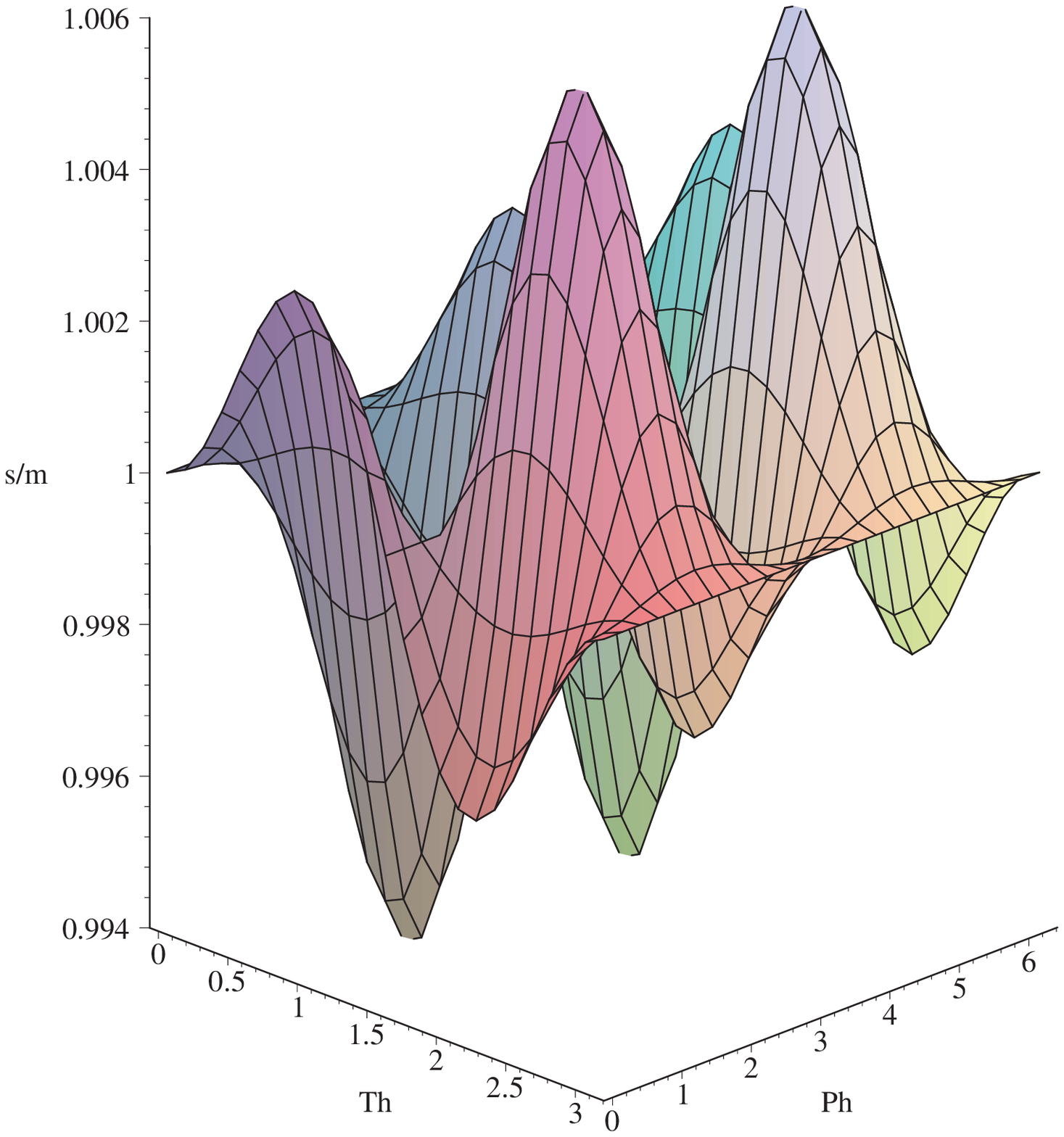}}}
\end{minipage}%
\hspace{0.5cm}
\begin{minipage}[t]{0.3 \textwidth}
\centering
\subfigure[\hspace{0.2cm} $a = M \, , \ O(\varepsilon^3)$]{
\label{fig:avg-Moller-Kerr-a=+1-e3}
\rotatebox{0}{\includegraphics[width = 5.0cm, height = 3.5cm, scale = 1]{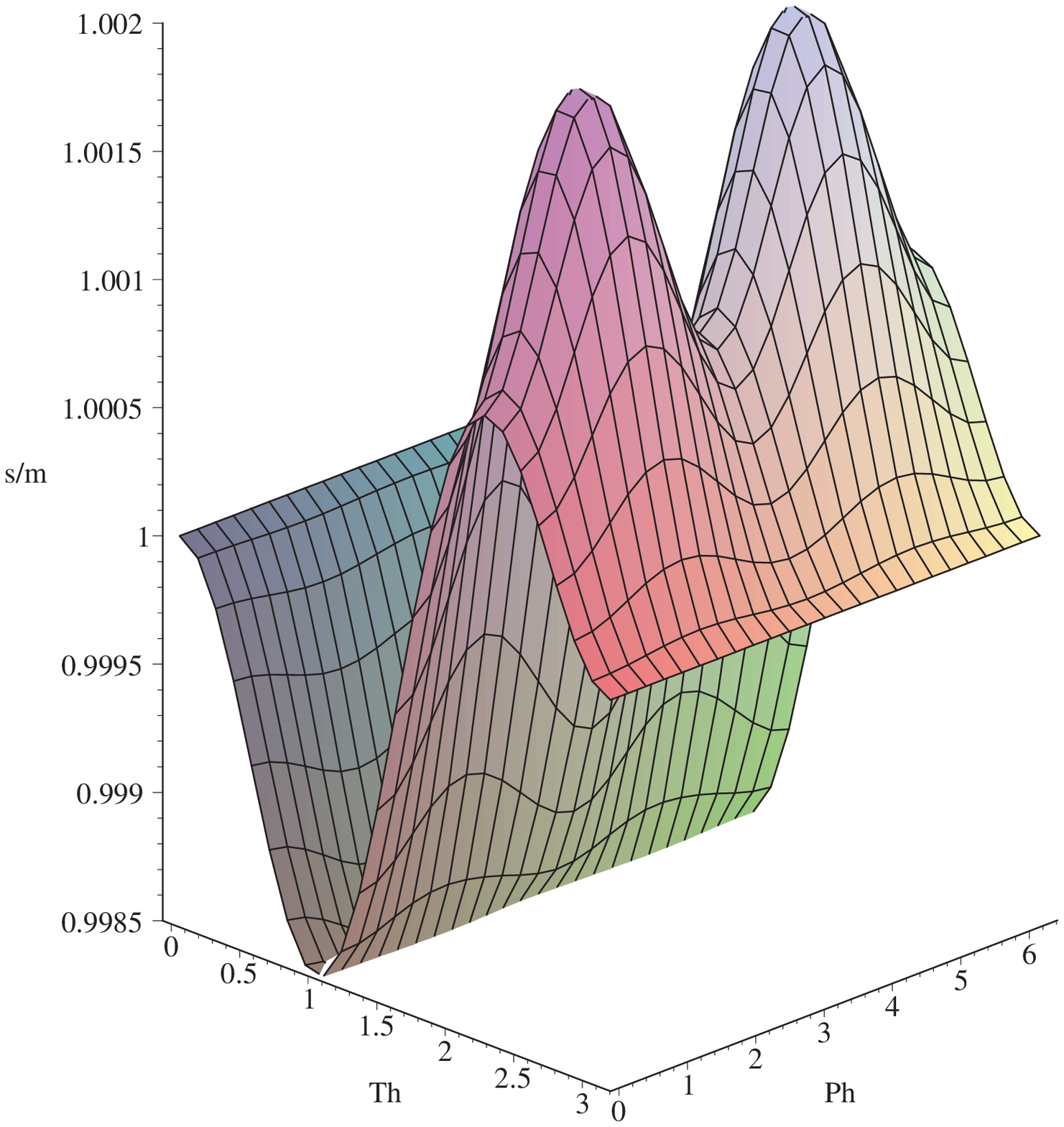}}}
\end{minipage}%
\hspace{0.5cm}
\begin{minipage}[t]{0.3 \textwidth}
\centering
\vspace{-3.5cm}
\caption{\label{fig:avg-Moller-Kerr-a=+1} Three-dimensional plot of the time-averaged M{\o}ller radius
$\lt\langle \rho \rt\rangle = \lt\langle s/m \rt\rangle$ as a function of $\hat{\th}$ and $\hat{\ph}$
for $r = 6 M$ and $a = M$.
Fig.~\ref{fig:avg-Moller-Kerr-a=+1-e2} shows a complicated peak and valley structure to $\lt\langle \rho \rt\rangle$
that simplifies somewhat in Fig.~\ref{fig:avg-Moller-Kerr-a=+1-e3}, with two peaks present.
\vspace{1mm}}
\end{minipage}
\end{figure*}

\begin{figure*}
\psfrag{Th}[cc][][1.8][0]{\Large $\hat{\th}$}
\psfrag{Ph}[cc][][1.8][0]{\Large $\hat{\ph}$}
\psfrag{s/m}[bc][][1.8][90]{\Large $\lt\langle s/m \rt\rangle \ [s_0/m_0]$}
\begin{minipage}[t]{0.3 \textwidth}
\centering
\subfigure[\hspace{0.2cm} $a = -M \, , \ O(\varepsilon^2)$]{
\label{fig:avg-Moller-Kerr-a=-1-e2}
\rotatebox{0}{\includegraphics[width = 5.0cm, height = 3.5cm, scale = 1]{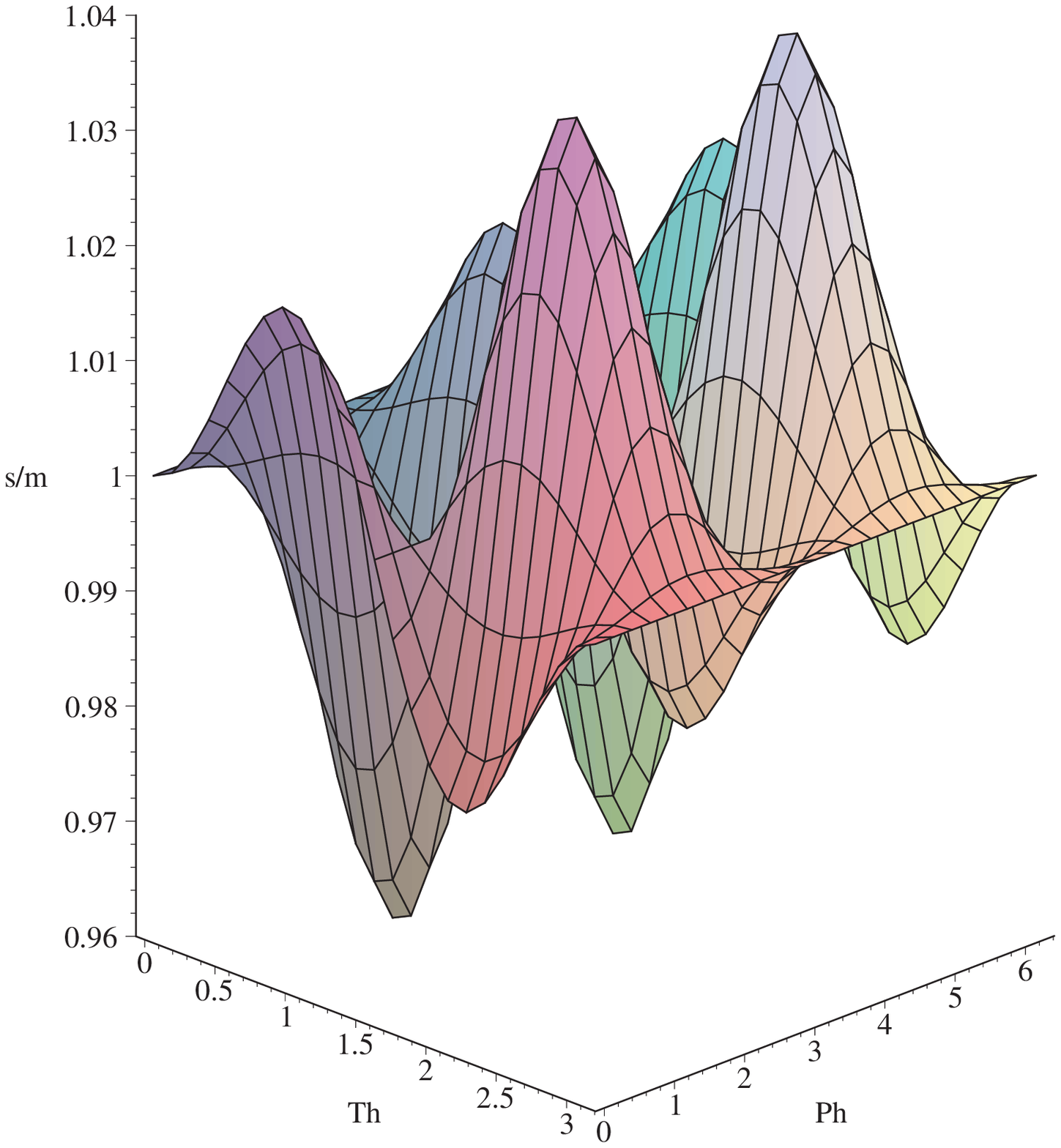}}}
\end{minipage}%
\hspace{0.5cm}
\begin{minipage}[t]{0.3 \textwidth}
\centering
\subfigure[\hspace{0.2cm} $a = -M \, , \ O(\varepsilon^3)$]{
\label{fig:avg-Moller-Kerr-a=-1-e3}
\rotatebox{0}{\includegraphics[width = 5.0cm, height = 3.5cm, scale = 1]{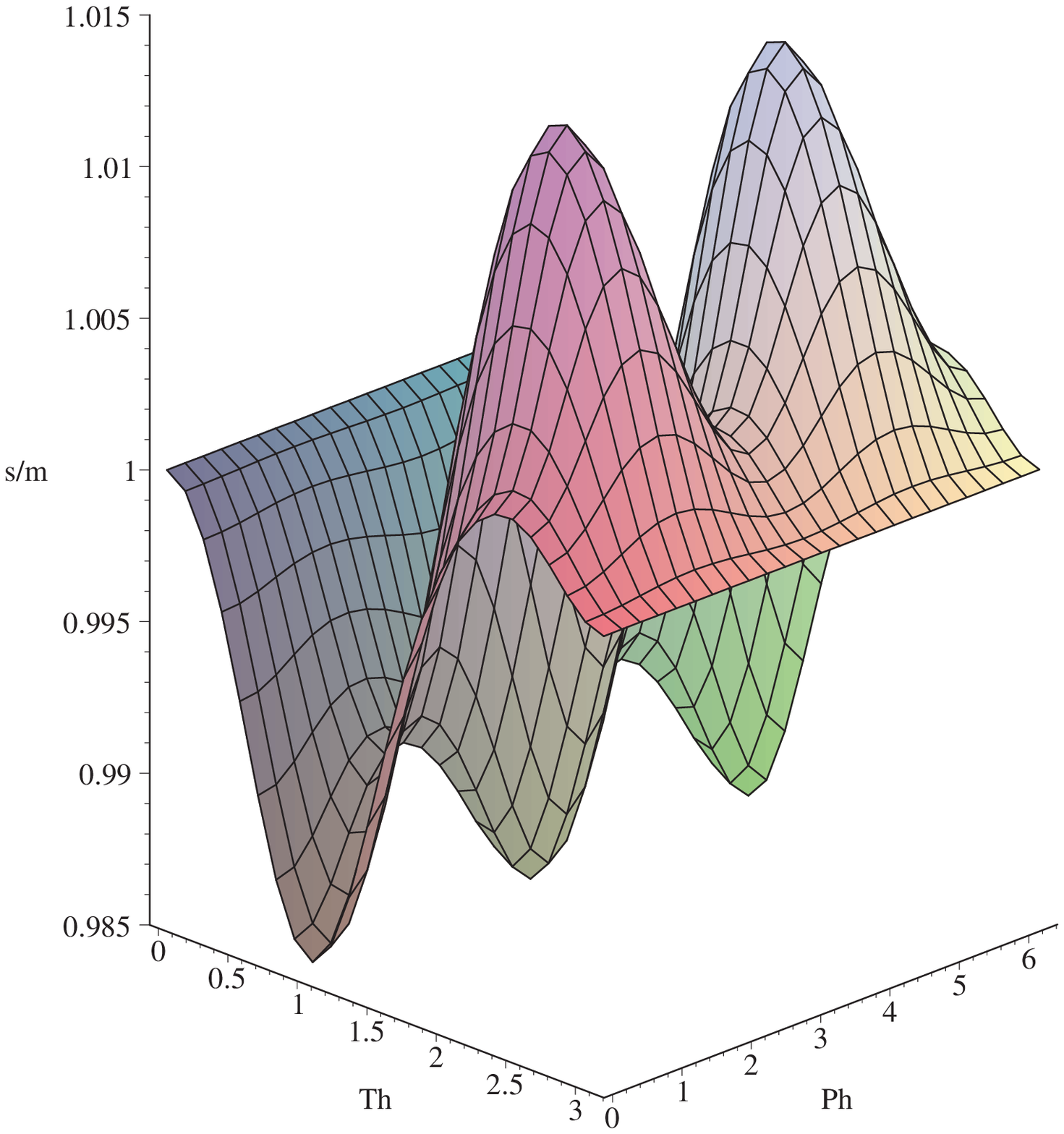}}}
\end{minipage}%
\hspace{0.5cm}
\begin{minipage}[t]{0.3 \textwidth}
\centering
\vspace{-3.5cm}
\caption{\label{fig:avg-Moller-Kerr-a=-1} Time-averaged M{\o}ller radius as a function of $\hat{\th}$ and $\hat{\ph}$
for $r = 6 M$ and $a = -M$.
It is clear that while the peak structure is essentially unchanged when compared to Figure~\ref{fig:avg-Moller-Kerr-a=+1},
the magnitude increases ten-fold when going from $a = M$ to $a = -M$.
\vspace{1mm}}
\end{minipage}
\end{figure*}

A further consideration involving the M{\o}ller radius is to examine its time-averaged value
$\lt\langle \rho \rt\rangle = \lt\langle s/m \rt\rangle$ as a function of the initial spin orientation angles $\hat{\th}$ and $\hat{\ph}$.
This leads to three-dimensional plots described by Figures~\ref{fig:avg-Moller-Kerr-a=+1} and \ref{fig:avg-Moller-Kerr-a=-1}
for $a = M$ and $a = -M$, respectively, which display expressions to both second- and third-order in $\varepsilon$ for
$\lt\langle \rho \rt\rangle$, and where $s_0/(m_0 \, r) = 10^{-1}$.
It is evident that all the plots reflect an even function symmetry with respect to $\hat{\ph} = \pi$.
According to Figs.~\ref{fig:avg-Moller-Kerr-a=+1-e2} and \ref{fig:avg-Moller-Kerr-a=-1-e2}, the $O(\varepsilon^2)$ expressions
lead to a non-trivial peak and valley structure in $\lt\langle \rho \rt\rangle$, while
Figs.~\ref{fig:avg-Moller-Kerr-a=+1-e3} and \ref{fig:avg-Moller-Kerr-a=-1-e3} for third-order in $\varepsilon$
effectively removes some of the structure for the region defined by $0 \leq \hat{\th} < \pi$, leaving two peaks to dominate.
While the shapes of the three-dimensional plots are effectively unchanged when comparing between
Figures~\ref{fig:avg-Moller-Kerr-a=+1} and \ref{fig:avg-Moller-Kerr-a=-1}, going from $a = M$ to $a = -M$ leads to a ten-fold increase
in magnitude, suggesting as expected that the counter-rotating black hole for $r = 6 M$ leads to a greater likelihood for
encountering instabilities within the spinning particle's orbit.

\begin{figure*}
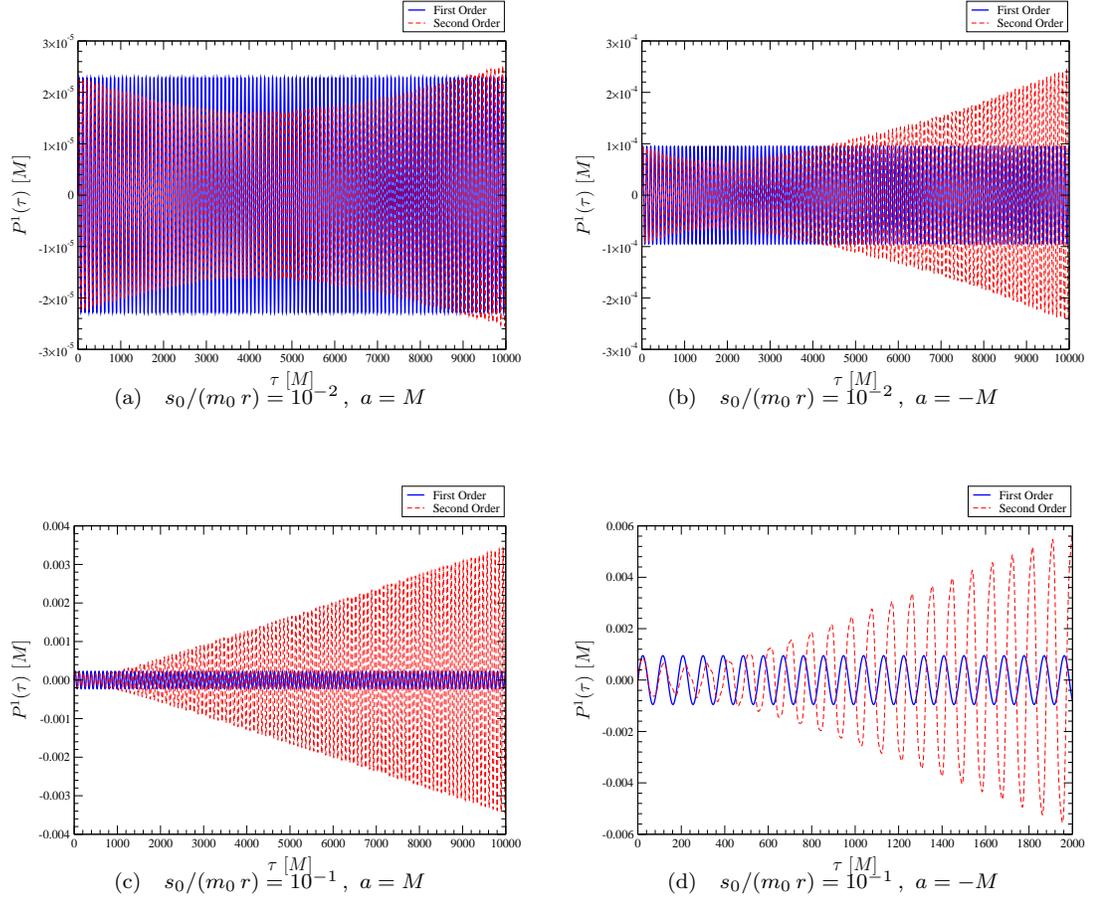

\psfrag{T}[tc][][1.8][0]{\Large $\tau \ [M]$}
\psfrag{P1}[bc][][1.8][0]{\Large $P^1(\tau) \ [M]$}
\begin{minipage}[t]{0.3 \textwidth}
\centering
\subfigure[\hspace{0.2cm} $s_0/(m_0 \, r) = 10^{-2} \, , \ a = M$]{
\label{fig:P1-Kerr-r=6-T=025-P=025-a=+1-mu=1e-2}
\rotatebox{0}{\includegraphics[width = 6.6cm, height = 5.0cm, scale = 1]{5a}}}
\end{minipage}%
\hspace{2.0cm}
\begin{minipage}[t]{0.3 \textwidth}
\centering
\subfigure[\hspace{0.2cm} $s_0/(m_0 \, r) = 10^{-2} \, , \ a = -M$]{
\label{fig:P1-Kerr-r=6-T=025-P=025-a=-1-mu=1e-2}
\rotatebox{0}{\includegraphics[width = 6.6cm, height = 5.0cm, scale = 1]{5b}}}
\end{minipage} \\
\vspace{0.8cm}
\begin{minipage}[t]{0.3 \textwidth}
\centering
\subfigure[\hspace{0.2cm} $s_0/(m_0 \, r) = 10^{-1} \, , \ a = M$]{
\label{fig:P1-Kerr-r=6-T=025-P=025-a=+1-mu=1e-1}
\rotatebox{0}{\includegraphics[width = 6.6cm, height = 5.0cm, scale = 1]{5c}}}
\end{minipage}%
\hspace{2.0cm}
\begin{minipage}[t]{0.3 \textwidth}
\centering
\subfigure[\hspace{0.2cm} $s_0/(m_0 \, r) = 10^{-1} \, , \ a = -M$]{
\label{fig:P1-Kerr-r=6-T=025-P=025-a=-1-mu=1e-1}
\rotatebox{0}{\includegraphics[width = 6.6cm, height = 5.0cm, scale = 1]{5d}}}
\end{minipage}
\caption{\label{fig:P1-Kerr-r=6-T=025-P=025} Radial component $P^1(\tau)$ of the linear momentum for
$r = 6 M$ and $\hat{\th} = \hat{\ph} = \pi/4$.
Fig.~\ref{fig:P1-Kerr-r=6-T=025-P=025-a=+1-mu=1e-2} shows a slightly growing amplitude due to the
second-order contribution in $\varepsilon$ for $s_0/(m_0 \, r) = 10^{-2}$, with a more moderate growth in
Fig.~\ref{fig:P1-Kerr-r=6-T=025-P=025-a=-1-mu=1e-2}.
The amplitude grows much more rapidly for
Figs.~\ref{fig:P1-Kerr-r=6-T=025-P=025-a=+1-mu=1e-1} and \ref{fig:P1-Kerr-r=6-T=025-P=025-a=-1-mu=1e-1} as
$s_0/(m_0 \, r) = 10^{-1}$.}
\end{figure*}

\begin{figure*}
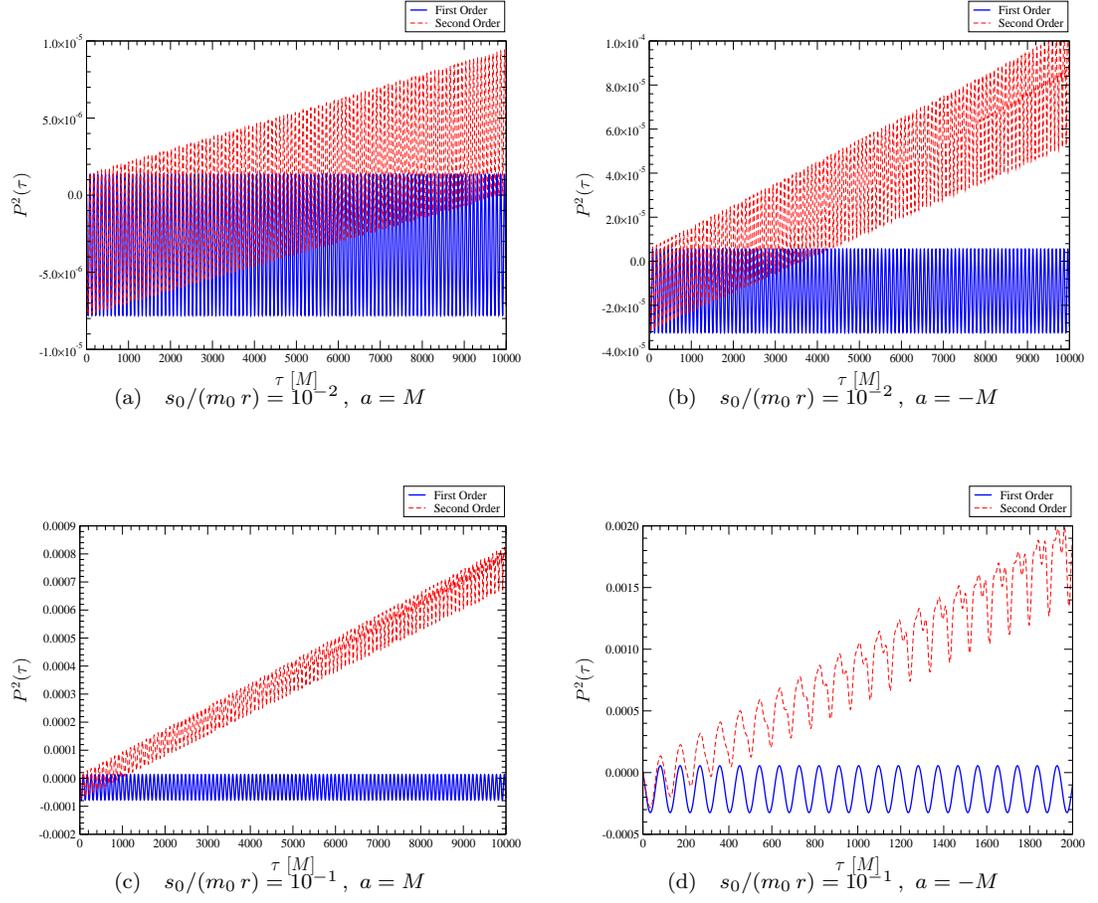

\psfrag{T}[tc][][1.8][0]{\Large $\tau \ [M]$}
\psfrag{P2}[bc][][1.8][0]{\Large $P^2(\tau)$}
\begin{minipage}[t]{0.3 \textwidth}
\centering
\subfigure[\hspace{0.2cm} $s_0/(m_0 \, r) = 10^{-2} \, , \ a = M$]{
\label{fig:P2-Kerr-r=6-T=025-P=025-a=+1-mu=1e-2}
\rotatebox{0}{\includegraphics[width = 6.6cm, height = 5.0cm, scale = 1]{6a}}}
\end{minipage}%
\hspace{2.0cm}
\begin{minipage}[t]{0.3 \textwidth}
\centering
\subfigure[\hspace{0.2cm} $s_0/(m_0 \, r) = 10^{-2} \, , \ a = -M$]{
\label{fig:P2-Kerr-r=6-T=025-P=025-a=-1-mu=1e-2}
\rotatebox{0}{\includegraphics[width = 6.6cm, height = 5.0cm, scale = 1]{6b}}}
\end{minipage} \\
\vspace{0.8cm}
\begin{minipage}[t]{0.3 \textwidth}
\centering
\subfigure[\hspace{0.2cm} $s_0/(m_0 \, r) = 10^{-1} \, , \ a = M$]{
\label{fig:P2-Kerr-r=6-T=025-P=025-a=+1-mu=1e-1}
\rotatebox{0}{\includegraphics[width = 6.6cm, height = 5.0cm, scale = 1]{6c}}}
\end{minipage}%
\hspace{2.0cm}
\begin{minipage}[t]{0.3 \textwidth}
\centering
\subfigure[\hspace{0.2cm} $s_0/(m_0 \, r) = 10^{-1} \, , \ a = -M$]{
\label{fig:P2-Kerr-r=6-T=025-P=025-a=-1-mu=1e-1}
\rotatebox{0}{\includegraphics[width = 6.6cm, height = 5.0cm, scale = 1]{6d}}}
\end{minipage}
\caption{\label{fig:P2-Kerr-r=6-T=025-P=025} Polar component $P^2(\tau)$ of the linear momentum for
$r = 6 M$ and $\hat{\th} = \hat{\ph} = \pi/4$.
The second-order contribution in $\varepsilon$ introduces a slight non-zero value in the net magnitude for
Figs.~\ref{fig:P2-Kerr-r=6-T=025-P=025-a=+1-mu=1e-2} and \ref{fig:P2-Kerr-r=6-T=025-P=025-a=-1-mu=1e-2}
with $s_0/(m_0 \, r) = 10^{-2}$, whose average slope becomes more pronounced for
Figs.~\ref{fig:P2-Kerr-r=6-T=025-P=025-a=+1-mu=1e-1} and \ref{fig:P2-Kerr-r=6-T=025-P=025-a=-1-mu=1e-1}
as $s_0/(m_0 \, r) = 10^{-1}$.}
\end{figure*}

\begin{figure*}
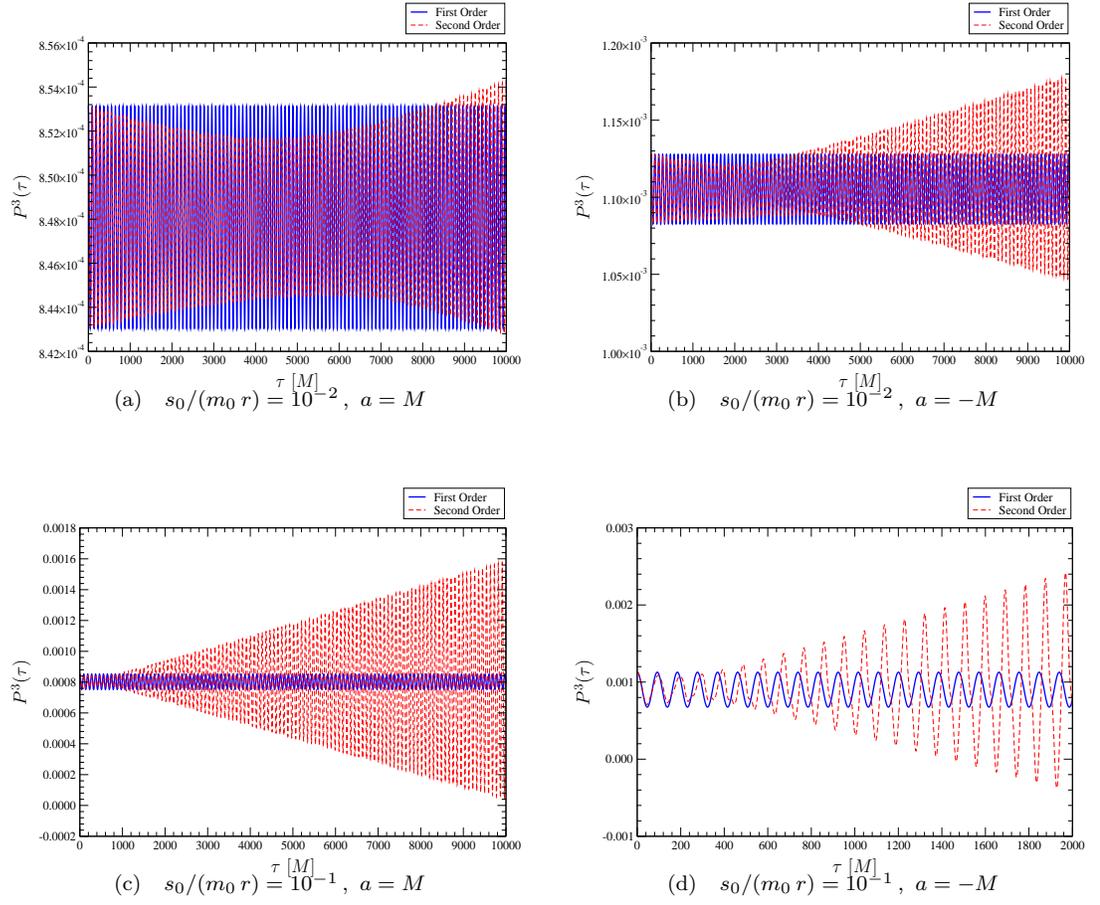

\psfrag{T}[tc][][1.8][0]{\Large $\tau \ [M]$}
\psfrag{P3}[bc][][1.8][0]{\Large $P^3(\tau)$}
\begin{minipage}[t]{0.3 \textwidth}
\centering
\subfigure[\hspace{0.2cm} $s_0/(m_0 \, r) = 10^{-2} \, , \ a = M$]{
\label{fig:P3-Kerr-r=6-T=025-P=025-a=+1-mu=1e-2}
\rotatebox{0}{\includegraphics[width = 6.6cm, height = 5.0cm, scale = 1]{7a}}}
\end{minipage}%
\hspace{2.0cm}
\begin{minipage}[t]{0.3 \textwidth}
\centering
\subfigure[\hspace{0.2cm} $s_0/(m_0 \, r) = 10^{-2} \, , \ a = -M$]{
\label{fig:P3-Kerr-r=6-T=025-P=025-a=-1-mu=1e-2}
\rotatebox{0}{\includegraphics[width = 6.6cm, height = 5.0cm, scale = 1]{7b}}}
\end{minipage} \\
\vspace{0.8cm}
\begin{minipage}[t]{0.3 \textwidth}
\centering
\subfigure[\hspace{0.2cm} $s_0/(m_0 \, r) = 10^{-1} \, , \ a = M$]{
\label{fig:P3-Kerr-r=6-T=025-P=025-a=+1-mu=1e-1}
\rotatebox{0}{\includegraphics[width = 6.6cm, height = 5.0cm, scale = 1]{7c}}}
\end{minipage}%
\hspace{2.0cm}
\begin{minipage}[t]{0.3 \textwidth}
\centering
\subfigure[\hspace{0.2cm} $s_0/(m_0 \, r) = 10^{-1} \, , \ a = -M$]{
\label{fig:P3-Kerr-r=6-T=025-P=025-a=-1-mu=1e-1}
\rotatebox{0}{\includegraphics[width = 6.6cm, height = 5.0cm, scale = 1]{7d}}}
\end{minipage}
\caption{\label{fig:P3-Kerr-r=6-T=025-P=025} Azimuthal component $P^3(\tau)$ of the linear momentum for
$r = 6 M$ and $\hat{\th} = \hat{\ph} = \pi/4$.
The second-order contribution in $\varepsilon$ introduces a slight non-zero change in the amplitude for
Fig.~\ref{fig:P3-Kerr-r=6-T=025-P=025-a=+1-mu=1e-2} with $s_0/(m_0 \, r) = 10^{-2}$, while
Fig.~\ref{fig:P3-Kerr-r=6-T=025-P=025-a=-1-mu=1e-2} shows a more moderate growth in amplitude.
This growth becomes strongly unbounded for Figs.~\ref{fig:P3-Kerr-r=6-T=025-P=025-a=+1-mu=1e-1} and
\ref{fig:P3-Kerr-r=6-T=025-P=025-a=-1-mu=1e-1} as $s_0/(m_0 \, r) = 10^{-1}$.}
\end{figure*}

\begin{figure*}
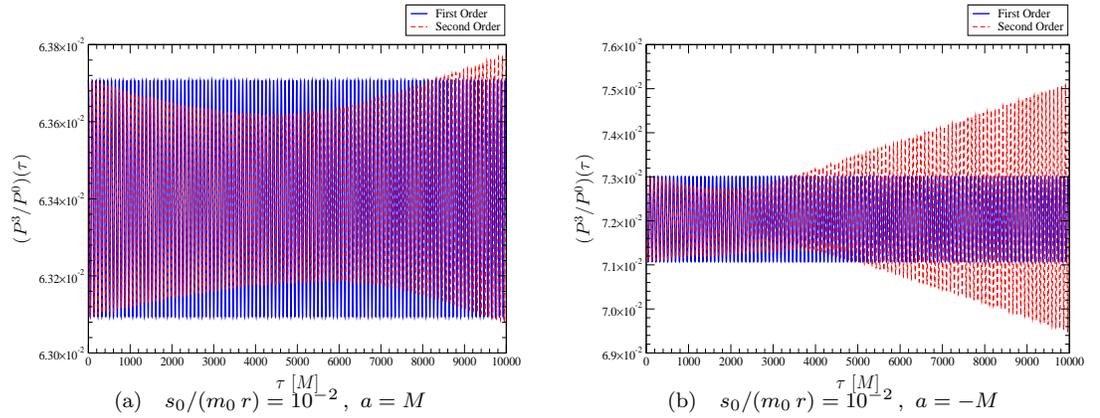

\psfrag{T}[tc][][1.8][0]{\Large $\tau \ [M]$}
\psfrag{P3/P0}[bc][][1.8][0]{\Large $(P^3/P^0)(\tau)$}
\begin{minipage}[t]{0.3 \textwidth}
\centering
\subfigure[\hspace{0.2cm} $s_0/(m_0 \, r) = 10^{-2} \, , \ a = M$]{
\label{fig:P3P0-Kerr-r=6-T=025-P=025-a=+1-mu=1e-2}
\rotatebox{0}{\includegraphics[width = 6.6cm, height = 5.0cm, scale = 1]{8a}}}
\end{minipage}%
\hspace{2.0cm}
\begin{minipage}[t]{0.3 \textwidth}
\centering
\subfigure[\hspace{0.2cm} $s_0/(m_0 \, r) = 10^{-2} \, , \ a = -M$]{
\label{fig:P3P0-Kerr-r=6-T=025-P=025-a=-1-mu=1e-2}
\rotatebox{0}{\includegraphics[width = 6.6cm, height = 5.0cm, scale = 1]{8b}}}
\end{minipage} \\
\hspace{2.0cm}
\caption{\label{fig:P3P0-Kerr-r=6-T=025-P=025} Ratio of $P^3(\tau)$ to $P^0(\tau)$ for $r = 6 M$ and $\hat{\th} = \hat{\ph} = \pi/4$.
When considering the expression to second-order in $\varepsilon$, it is evident from
Figs.~\ref{fig:P3P0-Kerr-r=6-T=025-P=025-a=+1-mu=1e-2} and \ref{fig:P3P0-Kerr-r=6-T=025-P=025-a=-1-mu=1e-2} that
the higher-order contribution becomes gradually unbounded compared to the constant ratio given by the first-order contribution in $\varepsilon$.}
\end{figure*}

It is useful to briefly examine the linear momentum components $P^\mu (\tau)$, given (\ref{P1-general}) and (\ref{P2-general}).
Figures~\ref{fig:P1-Kerr-r=6-T=025-P=025}--\ref{fig:P3-Kerr-r=6-T=025-P=025} display the radial, polar, and azimuthal
components of the linear momentum, while Figure~\ref{fig:P3P0-Kerr-r=6-T=025-P=025} displays the ratio
$P^3(\tau)/P^0(\tau)$, such as that described to first-order in $\varepsilon$ according to (\ref{P1-3-P1-0}).
For the radial component corresponding to $s_0/(m_0 \, r) = 10^{-2}$, Fig.~\ref{fig:P1-Kerr-r=6-T=025-P=025-a=+1-mu=1e-2}
for $a = M$ shows that the expression to second-order in $\varepsilon$ introduces a slight contraction in the amplitude of
$P^1(\tau)$ before expanding outwards.
This behaviour is also present in Fig.~\ref{fig:P1-Kerr-r=6-T=025-P=025-a=-1-mu=1e-2} for $a = -M$, though the outward growth is more pronounced,
the beginning of which roughly corresponds with the increase in $v(\tau)$ in Fig.~\ref{fig:v-Kerr-r=6-T=025-P=025-a=-1-mu=1e-2}
starting at $\tau = 3000 M$.
Not surprisingly, the $O(\varepsilon^2)$ expression becomes dominant in Figs.~\ref{fig:P1-Kerr-r=6-T=025-P=025-a=+1-mu=1e-1}
and \ref{fig:P1-Kerr-r=6-T=025-P=025-a=-1-mu=1e-1} when $s_0/(m_0 \, r) = 10^{-1}$,
which also corresponds with the respective increases in $v(\tau)$, as found in Figs.~\ref{fig:v-Kerr-r=6-T=025-P=025-a=+1-mu=1e-1}
and \ref{fig:v-Kerr-r=6-T=025-P=025-a=-1-mu=1e-1}.

The polar component corresponding to $s_0/(m_0 \, r) = 10^{-2}$ is described by Fig.~\ref{fig:P2-Kerr-r=6-T=025-P=025-a=+1-mu=1e-2}
for $a = M$ and Fig.~\ref{fig:P2-Kerr-r=6-T=025-P=025-a=-1-mu=1e-2} for $a = -M$, which indicate a slightly net positive
magnitude in $P^2(\tau)$ due to the expression to second-order in $\varepsilon$.
This outcome is somewhat surprising, since this suggests that the spinning particle will permanently leave the equatorial plane
under these conditions.
However, this may be more reflective of the choice for $r$, which hovers around the minimum value allowable for stable circular orbits.
Again, this outcome is more pronounced for the case of $s_0/(m_0 \, r) = 10^{-1}$, as shown in Figs.~\ref{fig:P2-Kerr-r=6-T=025-P=025-a=+1-mu=1e-1}
and \ref{fig:P2-Kerr-r=6-T=025-P=025-a=-1-mu=1e-1}.

For the azimuthal component corresponding to $s_0/(m_0 \, r) = 10^{-2}$, Figure~\ref{fig:P3-Kerr-r=6-T=025-P=025} behaves similarly to
that of Figure~\ref{fig:P1-Kerr-r=6-T=025-P=025}, particularly where it concerns Figs.~\ref{fig:P3-Kerr-r=6-T=025-P=025-a=+1-mu=1e-2}
and \ref{fig:P3-Kerr-r=6-T=025-P=025-a=-1-mu=1e-2} for $a = M$ and $a = -M$, respectively.
That is, the $O(\varepsilon^2)$ expression indicates a slight contraction in the amplitude of
$P^3(\tau)$ prior to an outward expansion.
Consistent with previous plots, Figs.~\ref{fig:P3-Kerr-r=6-T=025-P=025-a=+1-mu=1e-1} and \ref{fig:P3-Kerr-r=6-T=025-P=025-a=-1-mu=1e-1}
show a dominant growth of the amplitude of $P^3(\tau)$ for the expression to second-order in $\varepsilon$ and $s_0/(m_0 \, r) = 10^{-1}$.

Concerning the ratio $P^3(\tau)/P^0(\tau)$, this is presented in Figure~\ref{fig:P3P0-Kerr-r=6-T=025-P=025} for the case of
$s_0/(m_0 \, r) = 10^{-2}$, where Fig.~\ref{fig:P3P0-Kerr-r=6-T=025-P=025-a=+1-mu=1e-2} refers to $a = M$
and Fig.~\ref{fig:P3P0-Kerr-r=6-T=025-P=025-a=-1-mu=1e-2} corresponds to $a = -M$.
As expected, the first-order contribution in $\varepsilon$ leads to a constant ratio in $\tau$, consistent with (\ref{P1-3-P1-0}).
When the second-order contribution in $\varepsilon$ is added, the ratio exhibits a slight contraction followed by an outward
expansion, consistent with Figs.~\ref{fig:P3-Kerr-r=6-T=025-P=025-a=+1-mu=1e-2} and \ref{fig:P3-Kerr-r=6-T=025-P=025-a=-1-mu=1e-2}
for $P^3(\tau)$.
Nonetheless, it appears that the basic ratio remains constant for changing $\tau$.

\begin{figure*}
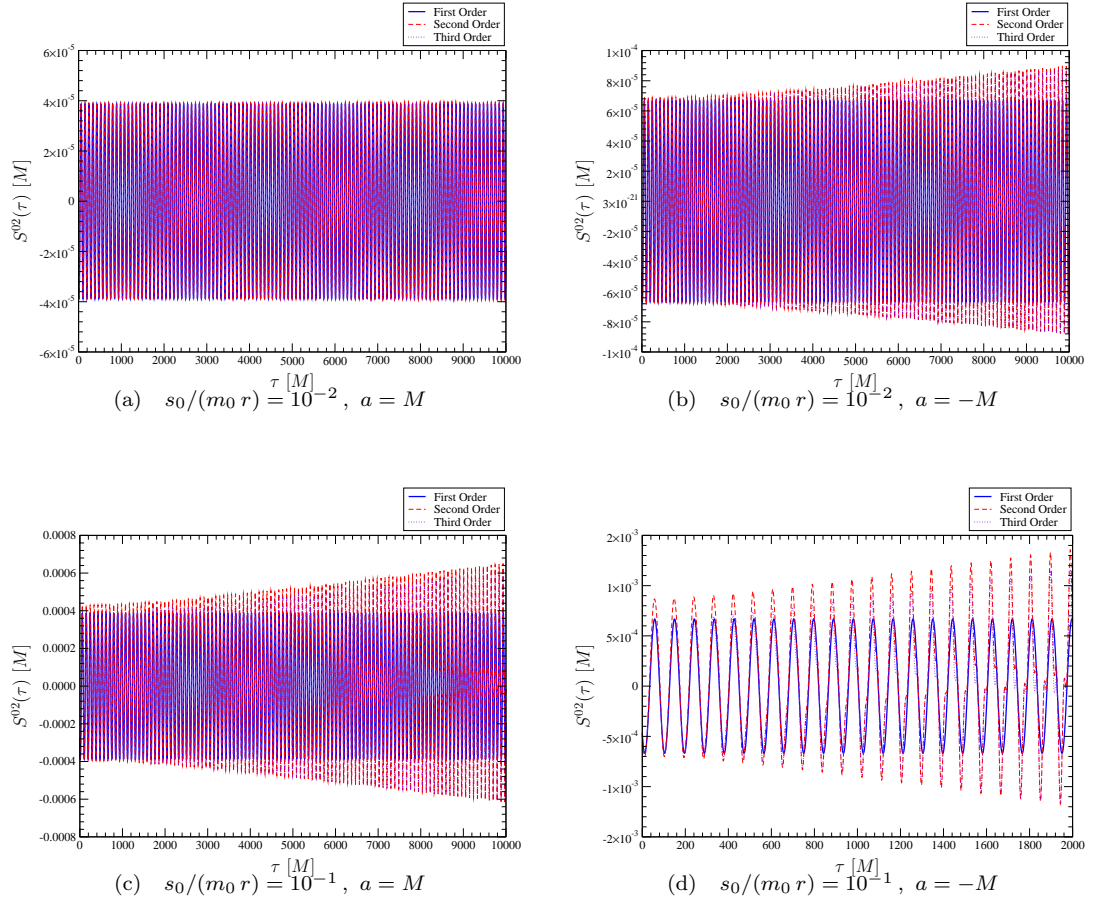

\psfrag{T}[tc][][1.8][0]{\Large $\tau \ [M]$}
\psfrag{S02}[bc][][1.8][0]{\Large $S^{02}(\tau) \ [M]$}
\begin{minipage}[t]{0.3 \textwidth}
\centering
\subfigure[\hspace{0.2cm} $s_0/(m_0 \, r) = 10^{-2} \, , \ a = M$]{
\label{fig:S02-Kerr-r=6-T=025-P=025-a=+1-mu=1e-2}
\rotatebox{0}{\includegraphics[width = 6.6cm, height = 5.0cm, scale = 1]{9a}}}
\end{minipage}%
\hspace{2.0cm}
\begin{minipage}[t]{0.3 \textwidth}
\centering
\subfigure[\hspace{0.2cm} $s_0/(m_0 \, r) = 10^{-2} \, , \ a = -M$]{
\label{fig:S02-Kerr-r=6-T=025-P=025-a=-1-mu=1e-2}
\rotatebox{0}{\includegraphics[width = 6.6cm, height = 5.0cm, scale = 1]{9b}}}
\end{minipage} \\
\vspace{0.8cm}
\begin{minipage}[t]{0.3 \textwidth}
\centering
\subfigure[\hspace{0.2cm} $s_0/(m_0 \, r) = 10^{-1} \, , \ a = M$]{
\label{fig:S02-Kerr-r=6-T=025-P=025-a=+1-mu=1e-1}
\rotatebox{0}{\includegraphics[width = 6.6cm, height = 5.0cm, scale = 1]{9c}}}
\end{minipage}%
\hspace{2.0cm}
\begin{minipage}[t]{0.3 \textwidth}
\centering
\subfigure[\hspace{0.2cm} $s_0/(m_0 \, r) = 10^{-1} \, , \ a = -M$]{
\label{fig:S02-Kerr-r=6-T=025-P=025-a=-1-mu=1e-1}
\rotatebox{0}{\includegraphics[width = 6.6cm, height = 5.0cm, scale = 1]{9d}}}
\end{minipage}
\caption{\label{fig:S02-Kerr-r=6-T=025-P=025} The $S^{02}(\tau)$ component of the spin tensor for
$r = 6 M$ and $\hat{\th} = \hat{\ph} = \pi/4$.
In Fig.~\ref{fig:S02-Kerr-r=6-T=025-P=025-a=+1-mu=1e-2}, there is virtually no contribution of the
second-order and third-order contributions in $\varepsilon$ to the amplitude, while there appears a slight
growth in amplitude for Fig.~\ref{fig:S02-Kerr-r=6-T=025-P=025-a=-1-mu=1e-2} due to these contributions.
In contrast, Figs.~\ref{fig:S02-Kerr-r=6-T=025-P=025-a=+1-mu=1e-1} and \ref{fig:S02-Kerr-r=6-T=025-P=025-a=-1-mu=1e-1}
show a noticeable increase in the amplitude as $s_0/(m_0 \, r) = 10^{-1}$ due to the higher-order contributions in $\varepsilon$.}
\end{figure*}

Finally, to explore the numerical properties of the spin tensor due to the generalized CMP approximation,
consider the example of $S^{02}(\tau)$, as presented in Figure~\ref{fig:S02-Kerr-r=6-T=025-P=025}, which
shows expressions up to third-order in $\varepsilon$.
It seems evident that the expression to second-order in $\varepsilon$ is dominant, as the
it is clear that the $O(\varepsilon^3)$ expression has no discernable impact on the amplitude.
According to Fig.~\ref{fig:S02-Kerr-r=6-T=025-P=025-a=+1-mu=1e-2} for $s_0/(m_0 \, r) = 10^{-2}$ and $a = M$,
the amplitude remains constant around $S^{02}(\tau) = 0$,
while Fig.~\ref{fig:S02-Kerr-r=6-T=025-P=025-a=-1-mu=1e-2} for $a = -M$ shows a gradual growth in the amplitude.
Consistent with previous plots, this effect becomes more pronounced in Figs.~\ref{fig:S02-Kerr-r=6-T=025-P=025-a=+1-mu=1e-1}
and \ref{fig:S02-Kerr-r=6-T=025-P=025-a=-1-mu=1e-1} for $s_0/(m_0 \, r) = 10^{-1}$.

\section{Conclusion}
\label{sec:6}

This paper outlines the generalization of an analytic perturbation approach to the Mathisson-Papapetrou-Dixon equations
for a spinning point particle, first introduced by Chicone, Mashhoon, and Punsly, with an application
to circular motion around a Kerr black hole.
The formalism shows the existence of ``radiative corrections'' to the particle's squared mass and spin magnitudes
due to spin-curvature interactions, represented in power series expansion form.
In performing the analysis, it is possible to semi-analytically identify the emergence of instabilities during the
particle's orbital motion, which serves as a basis for a more precise treatment in the future.

One of the underlying goals of the formalism presented in this paper is to determine the perturbed orbit
of the spinning particle according to the generalized CMP approximation, following the approach
taken earlier \cite{Mashhoon2}.
However, to do this properly requires a modification of the equations of motion to incorporate dissipative
effects due to gravitational radiation, which have not yet been taken into account.
Such a modification would most certainly require evaluation of the Teukolsky equations for determining
the radiation effects corresponding to an adiabatic inspiral for the spinning particle's orbit.
This is a non-trivial exercise with both conceptual and technical challenges to still overcome.
Once this is better understood, a determination of the perturbed orbit due to spin-curvature interactions
will be considered in a future publication.
For now, a second paper on the generalized CMP approximation in the Vaidya background is forthcoming \cite{Singh2} as a companion piece
to accompany and compare with this paper.

\begin{acknowledgments}
The author is thankful to Prof. Nader Mobed of the University of Regina for financial and moral support towards the
completion of this project.
\end{acknowledgments}

\begin{appendix}

\section{Fermi-Frame Riemann Tensor Components}
\label{appendix:riemann-frame-components}
\renewcommand{\theequation}{A.\arabic{equation}}
\setcounter{subsection}{0}
\setcounter{equation}{0}

Given that the nonzero Riemann curvature tensor components in standard Boyer-Lindquist co-ordinates are
\begin{widetext}
\be
R_{0101} & = & -{M r \over \Sg^3 \Dl} \lt(\Sg - 4 a^2 \, \cos^2 \th\rt)\lt(2 \Dl + a^2 \, \sin^2 \th \rt),
\nl
R_{0102} & = & {3 M a^2 \over \Sg^3} \lt(4 r^2 - \Sg\rt) \sin \th \, \cos \th \, ,
\nl
R_{0113} & = & -{M a r \over \Sg^3 \Dl} \lt(4 r^2 - 3 \Sg\rt)\lt(r^2 + a^2 + 2 \Dl \rt) \sin^2 \th \, ,
\nl
R_{0123} & = & {M a \over \Sg^3} \lt(4 r^2 - \Sg\rt)\lt(2 \Sg + 3 a^2 \, \sin^2 \th\rt) \sin \th \, \cos \th \, ,
\nl
R_{0202} & = & {M r \over \Sg^3} \lt(4 r^2 - 3 \Sg\rt)\lt(\Dl + 2 a^2 \, \sin^2 \th\rt),
\nl
R_{0213} & = & {M a \over \Sg^3} \lt(4 r^2 - \Sg\rt)\lt(\Sg + 3 a^2 \, \sin^2 \th \rt) \sin \th \, \cos \th \, ,
\nl
R_{0223} & = & {M a r \over \Sg^3} \lt(4 r^2 - 3 \Sg\rt)\lt[2\lt(r^2 + a^2\rt) + \Dl \rt]\sin^2 \th \, ,
\nl
R_{0303} & = & {M r \Dl \over \Sg^3} \lt(4 r^2 - 3 \Sg\rt) \sin^2 \th \, ,
\nl
R_{0312} & = & -{M a \over \Sg^2} \lt(4 r^2 - \Sg\rt) \sin \th \, \cos \th \, ,
\nl
R_{1212} & = & -{M r \over \Sg \Dl} \lt(4 r^2 - 3 \Sg\rt) \, ,
\nl
R_{1313} & = & -{M r \over \Sg^3 \Dl} \lt(4 r^2 - 3 \Sg\rt) \lt[\lt(r^2 + a^2\rt)^2 + 2 a^2 \Dl \, \sin ^2 \th \rt] \sin^2 \th \, ,
\nl
R_{1323} & = & {3 M a^2 \over \Sg^3} \lt(4 r^2 - \Sg\rt) \lt(r^2 + a^2\rt) \sin^3 \th \, \cos \th \, ,
\nl
R_{2323} & = & {M r \over \Sg^3} \lt(4 r^2 - 3 \Sg\rt) \lt[2\lt(r^2 + a^2\rt)^2 + a^2 \Dl \, \sin ^2 \th \rt] \sin^2 \th \, ,
\ee
\end{widetext}
where
\be
\Sg & = & r^2 + a^2 \, \cos^2 \th \, ,
\label{Sg=}
\nl
\Dl & = & r^2 + a^2 - 2 \, M \, r \, ,
\label{Dl=}
\ee
the nonzero components of the Riemann curvature tensor ${}^F{}R_{\Cmu \Cnu \Cal \Cbt}$ in the Fermi frame are listed as follows:
\begin{widetext}
\be
{}^F{}R_{\ze \on \ze \on} & = &
{\Dl \over N^2 \, r^2} \lt[\lt(1 + a \, \OmK\rt)^2 R_{0101} - {2 \, \OmK \over \sin \th} \lt(1 + a \, \OmK\rt) R_{0113}
+ {\OmK^2 \over \sin^2 \th} \, R_{1313}\rt] \cos^2 \lt(\OmK \, \tau\rt)
\nn
\nn
\nn
&  &{} + {1 \over N^2 \, \Dl \sin^2 \th} \lt[E + \OmK \lt(a \, E - L\rt) \rt]^2 R_{0303} \, \sin^2 \lt(\OmK \, \tau\rt) \, ,
\nl
\nn
\nn
{}^F{}R_{\ze \on \ze \tw} & = &
{\sqrt{\Dl} \over N^2 \, r^2} \lt[\lt(1 + a \, \OmK\rt)^2 R_{0102} - {\OmK \over \sin \th} \lt(1 + a \, \OmK\rt) \lt(R_{0213} + R_{0123}\rt)
+ {\OmK^2 \over \sin^2 \th} \, R_{1323}\rt] \cos \lt(\OmK \, \tau\rt) \, ,
\nl
\nn
\nn
{}^F{}R_{\ze \on \ze \tr} & = &
{1 \over N^2 \, \sin \th} \lt\{
{\Dl \over r^2} \lt[\lt(1 + a \, \OmK\rt) \lt[\lt(1 + a \, \OmK\rt) \sin \th \, R_{0101} - 2 \, \OmK \, R_{0113} \rt]
+ {\OmK^2 \over \sin \th} \, R_{1313} \rt] \rt.
\nn
\nn
\nn
&  &{} - \lt. {1 \over \Dl \, \sin \th} \, \lt[E + \OmK \lt(a \, E - L\rt) \rt]^2 R_{0303} \rt\} \sin \lt(\OmK \, \tau \rt) \, \cos \lt(\OmK \, \tau\rt) \, ,
\nl
\nn
\nn
{}^F{}R_{\ze \on \on \tw} & = &
{1 \over N \, r^2} \lt\{
\lt(1 + a \, \OmK\rt) \lt({E  \over \sin \th} \, R_{0123} - L \, R_{0102} \rt)
- {\OmK \over \sin \th} \lt({E  \over \sin \th} \, R_{1323} - L \, R_{0213} \rt) \rt.
\nn
\nn
\nn
&  &{} - \lt. {1 \over \sin \th} \, \lt[E + \OmK \lt(a \, E - L\rt) \rt] R_{0312} \rt\} \sin \lt(\OmK \, \tau \rt) \, \cos \lt(\OmK \, \tau\rt) \, ,
\nl
\nn
\nn
{}^F{}R_{\ze \on \on \tr} & = &
{\sqrt{\Dl} \over N \, r^2} \lt\{{1 \over \sin \th} \, \lt[E + \OmK \lt(a \, E + L\rt) \rt] R_{0113}
- \lt[{\OmK \, E  \over \sin^2 \th} \, R_{1313} + \lt(1 + a \, \OmK\rt) L \, R_{0101} \rt] \rt\} \cos \lt(\OmK \, \tau\rt) \, ,
\nl
\nn
\nn
{}^F{}R_{\ze \on \tw \tr} & = &
{1 \over N \, r^2} \lt\{\lt[
\lt(1 + a \, \OmK\rt) \lt({E  \over \sin \th} \, R_{0123} - L \, R_{0102} \rt)
- {\OmK \over \sin \th} \lt({E  \over \sin \th} \, R_{1323} - L \, R_{0213} \rt) \rt] \cos^2 \lt(\OmK \, \tau\rt) \rt.
\nn
\nn
\nn
&  &{} + \lt. {1 \over \sin \th} \, \lt[E + \OmK \lt(a \, E - L\rt) \rt] R_{0312} \, \sin^2 \lt(\OmK \, \tau \rt) \rt\}   \, ,
\nl
\nn
\nn
{}^F{}R_{\ze \tw \ze \tw} & = &
{1 \over N^2 \, r^2} \lt[\lt(1 + a \, \OmK\rt)^2 R_{0202} - {2 \, \OmK \over \sin \th} \lt(1 + a \, \OmK\rt) R_{0223} + {\OmK^2 \over \sin^2 \th} \, R_{2323} \rt] \, ,
\nl
\nn
\nn
{}^F{}R_{\ze \tw \ze \tr} & = &
{\sqrt{\Dl} \over N^2 \, r^2} \lt\{\lt(1 + a \, \OmK\rt) \lt[
\lt(1 + a \, \OmK\rt) R_{0102} - {\OmK \over \sin \th} \lt(R_{0123} + R_{0213}\rt)\rt] + {\OmK^2 \over \sin^2 \th} \, R_{1323} \rt\}
\sin \lt(\OmK \, \tau\rt) \, ,
\nl
\nn
\nn
{}^F{}R_{\ze \tw \on \tw} & = &
{1 \over N \, \sqrt{\Dl} \, r^2} \lt\{{1 \over \sin \th} \, \lt[E + \OmK \lt(a \, E + L\rt) \rt] R_{0223} -
\lt[{\OmK \, E \over \sin^2 \th} \, R_{2323} + \lt(1 + a \, \OmK\rt) L \, R_{0202} \rt] \rt\} \sin \lt(\OmK \, \tau\rt) \, ,
\nl
\nn
\nn
{}^F{}R_{\ze \tw \on \tr} & = &
{1 \over N \, r^2} \lt[
\lt(1 + a \, \OmK\rt) \lt({E \over \sin \th} \, R_{0213} - L \, R_{0102} \rt)
- {\OmK \over \sin \th} \lt({E \over \sin \th} \, R_{1323} - L \, R_{0123}\rt) \rt] \, ,
\ee

\be
{}^F{}R_{\ze \tw \tw \tr} & = &
{1 \over N \, \sqrt{\Dl} \, r^2} \lt\{ {1 \over \sin \th} \, \lt[E + \OmK \lt(a \, E + L\rt) \rt] R_{0223}
- \lt[{\OmK \, E \over \sin^2 \th} \, R_{2323} + \lt(1 + a \, \OmK\rt) L \, R_{0202} \rt] \rt\} \cos \lt(\OmK \, \tau\rt) \, ,
\nl
\nn
\nn
{}^F{}R_{\ze \tr \ze \tr} & = &
{\Dl \over N^2 \, r^2} \lt[\lt(1 + a \, \OmK\rt)^2 R_{0101} - {2 \, \OmK \over \sin \th} \lt(1 + a \, \OmK\rt) R_{0113}
+ {\OmK^2 \over \sin^2 \th} \, R_{1313}\rt] \sin^2 \lt(\OmK \, \tau\rt)
\nn
\nn
\nn
&  &{} + {1 \over N^2 \, \Dl \sin^2 \th} \lt[E + \OmK \lt(a \, E - L\rt) \rt]^2 R_{0303} \, \cos^2 \lt(\OmK \, \tau\rt) \, ,
\nl
\nn
\nn
{}^F{}R_{\ze \tr \on \tw} & = &
{1 \over N \, r^2} \lt\{
\lt[\lt(1 + a \, \OmK\rt) \lt({E  \over \sin \th} \, R_{0123} - L \, R_{0102} \rt)
- {\OmK \over \sin \th} \lt({E  \over \sin \th} \, R_{1323} - L \, R_{0213} \rt) \rt] \sin^2 \lt(\OmK \, \tau\rt) \rt.
\nn
\nn
\nn
&  &{} + \lt. {1 \over \sin \th} \, \lt[E + \OmK \lt(a \, E - L\rt) \rt] R_{0312} \, \cos^2 \lt(\OmK \, \tau\rt) \rt\}  \, ,
\nl
\nn
\nn
{}^F{}R_{\ze \tr \on \tr} & = &
{\sqrt{\Dl} \over N \, r^2} \lt\{{1 \over \sin \th} \, \lt[E + \OmK \lt(a \, E + L\rt) \rt] R_{0113}
- \lt[{\OmK \, E  \over \sin^2 \th} \, R_{1313} + \lt(1 + a \, \OmK\rt) L \, R_{0101} \rt] \rt\} \sin \lt(\OmK \, \tau\rt) \, ,
\nl
\nn
\nn
{}^F{}R_{\ze \tr \tw \tr} & = &
{1 \over N \, r^2} \lt\{
\lt(1 + a \, \OmK\rt) \lt({E  \over \sin \th} \, R_{0123} - L \, R_{0102} \rt)
- {\OmK \over \sin \th} \lt({E  \over \sin \th} \, R_{1323} - L \, R_{0213} \rt) \rt.
\nn
\nn
\nn
&  &{} - \lt. {1 \over \sin \th} \, \lt[E + \OmK \lt(a \, E - L\rt) \rt] R_{0312} \rt\} \sin \lt(\OmK \, \tau \rt) \, \cos \lt(\OmK \, \tau\rt) \, ,
\nl
\nn
\nn
{}^F{}R_{\on \tw \on \tw} & = &
{1 \over \Dl \, r^2} \lt[{E \over \sin \th} \lt({E \over \sin \th} \, R_{2323} - 2 \, L \, R_{0223} \rt) + L^2 \, R_{0202} \rt] \sin^2 \lt(\OmK \, \tau\rt)
+ {\Dl \over r^4} \, R_{1212} \, \cos^2 \lt(\OmK \, \tau\rt) \, ,
\nl
\nn
\nn
{}^F{}R_{\on \tw \on \tr} & = &
{1 \over \sqrt{\Dl} \, r^2} \lt[{E \over \sin \th} \lt({E \over \sin \th} \, R_{1323} - L \, R_{0123} \rt)
- L \lt({E \over \sin \th} \, R_{0213} - L \, R_{0102} \rt) \rt] \sin^2 \lt(\OmK \, \tau\rt) \, ,
\nl
\nn
\nn
{}^F{}R_{\on \tw \tw \tr} & = & \lt\{
{1 \over \Dl \, r^2} \lt[{E \over \sin \th} \lt({E \over \sin \th} \, R_{2323} - 2 \, L \, R_{0223} \rt) + L^2 \, R_{0202} \rt]
- {\Dl \over r^4} \, R_{1212} \rt\} \sin \lt(\OmK \, \tau\rt) \, \cos \lt(\OmK \, \tau\rt) \, ,
\nl
\nn
\nn
{}^F{}R_{\on \tr \on \tr} & = &
{1 \over r^2} \lt[{E \over \sin \th} \lt({E \over \sin \th} \, R_{1313} - 2 \, L \, R_{0113} \rt) + L^2 \, R_{0101} \rt] \, ,
\nl
\nn
\nn
{}^F{}R_{\on \tr \tw \tr} & = &
{1 \over \sqrt{\Dl} \, r^2} \lt\{ {E \over \sin \th} \lt[{E \over \sin \th} \, R_{1323} - L \, \lt(R_{0123} + R_{0213} \rt) \rt] + L^2 \, R_{0102} \rt\}
\cos \lt(\OmK \, \tau\rt) \, ,
\nl
\nn
\nn
{}^F{}R_{\tw \tr \tw \tr} & = &
{1 \over \Dl \, r^2} \lt[{E \over \sin \th} \lt({E \over \sin \th} \, R_{2323} - 2 \, L \, R_{0223} \rt) + L^2 \, R_{0202} \rt] \cos^2 \lt(\OmK \, \tau\rt)
+ {\Dl \over r^4} \, R_{1212} \, \sin^2 \lt(\OmK \, \tau\rt) \, .
\ee
\end{widetext}

\newpage

\section{Selected Plots for the Generalized CMP Approximation of the MPD Equations}
\label{appendix:plots}
\renewcommand{\theequation}{B.\arabic{equation}}
\setcounter{subsection}{0}
\setcounter{equation}{0}

\end{appendix}

\end{document}